\documentclass[conference]{IEEEtran}

\ifCLASSINFOpdf
  \usepackage[pdftex]{graphicx}
\else
\fi

\usepackage{url}
\usepackage{color,soul}
\usepackage{booktabs}
\usepackage{multirow}

\usepackage{algorithm}
\usepackage{listings}
\usepackage[noend]{algpseudocode}
\usepackage{sansmath}
\usepackage{xurl}
\usepackage{balance}
\usepackage{longtable}
\usepackage{makecell}
\usepackage{pseudo}
\usepackage{xcolor}
\usepackage[listings]{tcolorbox}
\tcbuselibrary{listings,theorems}

\definecolor{gainsboro}{rgb}{0.86, 0.86, 0.86}

\usepackage{amsmath}
\usepackage{filecontents}
\usepackage{caption}
\usepackage{subcaption}

\usepackage{tikz}

\newcommand{\our}{\textsc{CloneBuster}}

\newcommand{\catA}{{FIm}}
\newcommand{\catB}{{ForKVS}}
\newcommand{\catC}{{BUG}}

\usepackage[]{graphicx}

\begin{document}

\title{The Real Menace of Cloning Attacks on SGX Applications}

\author{\IEEEauthorblockN{Samira Briongos}
\IEEEauthorblockA{NEC Laboratories Europe\\
samira.briongos@neclab.eu}
\and
\IEEEauthorblockN{Ghassan Karame}
\IEEEauthorblockA{Ruhr University Bochum\\
ghassan@karame.org}
\and
\IEEEauthorblockN{Claudio Soriente}
\IEEEauthorblockA{NEC Laboratories Europe\\
claudio.soriente@neclab.eu}
\and
\IEEEauthorblockN{Annika Wilde}
\IEEEauthorblockA{Ruhr University Bochum\\
annika.wilde@rub.de}
}

\maketitle

\begin{abstract}
Trusted Execution Environments (TEEs) are gaining popularity as an effective means to provide confidentiality in the cloud. TEEs, such as Intel SGX, suffer from so-called rollback and cloning attacks (often referred to as forking attacks).  Rollback attacks are enabled by the lack of freshness guarantees for sealed data; cloning attacks stem from the inability to determine if other instances of an enclave are running on the same platform. While rollback attacks have been extensively studied by the community, cloning attacks have been, unfortunately, less investigated. To address this gap, we extensively study and thoroughly analyze the susceptibility of 72 SGX-based proposals to cloning attacks. Our results show that roughly 20\% of the analyzed proposals are insecure against cloning attacks---including those applications that rely on monotonic counters and are, therefore, secure against rollback attacks.
\end{abstract}

\section{Introduction}

Trusted Execution Environments (TEE), such as Intel SGX, enable user processes to run in isolation (i.e., in so-called enclaves) from other software on the same platform, including the OS. 
TEEs are gaining popularity as an effective means to provide confidentiality in the cloud, a rapidly growing market.
Intel SGX applications are, however, susceptible to so-called \emph{forking attacks}, where the adversary partitions the set of clients and provides them with different views of the system. Forking attacks may be mounted either by cloning an enclave or by rolling back its state~\cite{brandenburger17dsn}. Rollback attacks exploit the fact that the sealing functionality of Intel SGX lacks freshness guarantees. This opens the door for a malicious OS to feed a victim enclave with a stale state whenever the enclave requests to unseal its state from storage---thereby ``rolling back'' the enclave to a previous state.
Cloning attacks leverage the fact that Intel SGX does not provide means to control the number of enclaves with the same binary that a malicious OS can launch on the same machine. We identify two important gaps concerning forking attacks:

\begin{itemize}
    \item \textbf{Lack of awareness on the impact of cloning attacks:} In current SGX-based applications, cloning attacks seem to be largely dismissed. While rollback attacks are typically well discussed and addressed by means of monotonic counters, the integration of forking mitigations is often not discussed in detail---opening an attack window for cloning if not implemented properly and in the right places.
    \item \textbf{No practical forking mitigations:} To date, several solutions to thwart cloning attacks have been proposed. Current forking mitigations rely on a centralized trusted third party (TTP)~\cite{DBLP:conf/ccs/DijkRSD07,strackx16usenix} such as a monotonic counter, or a distributed one~\cite{DBLP:conf/ndss/Kaptchuk0M19,brandenburger17dsn,ROTE2017,NARRATOR2022}. These solutions are largely impractical---requiring monotonic counters, which are unavailable for SGX, or relying on TTPs, which are hard to find in practice. Further, the solution itself may require protection against cloning.
\end{itemize}

Forking attacks against enclaves---either by rollback or by cloning---result in serious consequences in a number of applications ranging from digital payments~\cite{lind2017teechain} to password-based authentication~\cite{strackx16usenix}. For example, in a password manager application, forking attacks may allow an adversary to brute-force a password despite rate-limiting measures adopted by the application. Similarly, in a payment application, an adversary could spend the same coins in multiple payments by reverting the state of its account balance.

A comprehensive solution to thwart forking attacks requires a centralized trusted third party (TTP)~\cite{DBLP:conf/ccs/DijkRSD07} or a distributed one~\cite{DBLP:conf/ndss/Kaptchuk0M19,brandenburger17dsn,ROTE2017,NARRATOR2022}. Unfortunately, in most real-world applications, TTPs are hard to find. Moreover, some TTP-based solutions might themselves be subject to cloning attacks during the initialization process, unless the initialization involves yet another TTP (e.g., a trusted administrator~\cite{ROTE2017} or a blockchain~\cite{NARRATOR2022}). Without TTPs, most applications can mitigate forking attacks based on rollbacks by means of hardware-based monotonic counters~\cite{strackx16usenix}. However, an application that uses monotonic counters can still be cloned---making it still susceptible to forking attacks.
Moreover, hardware monotonic counters are not available for SGX in the cloud.

Our main aim in this paper is to raise the awareness of the detrimental impact of cloning attacks on SGX applications. We show that such attacks are straightforward to realize and not trivial to defend. To this end, we extensively analyze the security of 72 SGX-based applications against rollback and cloning attacks. Our findings show that 14 of those applications (i.e., roughly 20\%) are vulnerable to forking attacks based on cloning (cf. Section~\ref{sec:analysis}). Among these applications, we identify three categories of cloning attacks applicable to all vulnerable applications (cf. Section~\ref{sec:attacks}). 

This work extends our previous ACSAC'23 paper~\cite{clonebuster_acsac}. Our main new contribution in this work, compared to our previous conference paper, is a detailed analysis of the 72 SGX-based applications that we consider. Namely, as a result of a semi-manual and thorough analysis of the applications, we identify three categories of cloning-based attacks and describe each category in detail.

\section{Background}
\label{sec:background}

\subsection{Intel SGX}
Intel Software Guard Extensions (SGX) is an x86 instruction set extension that offers hardware-based isolation to trusted applications that run in so-called \emph{enclaves}~\cite{IntelArchitecture}.
Enclave isolation leverages dedicated, hardware-protected memory called Enclave Page Cache (EPC). The OS is responsible for loading the enclave's software in the EPC while the processor keeps track of deployed enclaves and their memory pages in the Enclave Page Cache Map (EPCM). By leveraging EPCM, the hardware restricts access to EPC from processes running at higher privilege levels, including the OS or the hypervisor. In particular, the Memory Management Unit (MMU) uses the EPCM to abort any enclave memory access attempt that has not been issued by the enclave itself or that does not comply with the specified read/write/execute permissions.

Attestation allows the platform to issue publicly verifiable statements of an enclave's software configuration. In particular, each application enclave has two identities: one called MRENCLAVE computed as the hash of the enclave binary loaded into memory; the other identity is called MRSIGNER and identifies the enclave developer. During attestation, a designated system enclave outputs a signature over both identities to certify that the application runs in an enclave on an SGX-enabled platform.

Intel SGX also allows enclaves to store encrypted data on disk. This is achieved via a ``sealing'' interface that uses hardware-managed cryptographic keys. Sealed data is encrypted and authenticated using keys that are dependent on the platform and on one of the enclave identities. Sealing data against the MRENCLAVE identity ensures that only enclaves loaded with the same binary on the same platform can unseal it; on the other hand, sealing data against the MRSIGNER identity ensures that all enclaves running on the same platform and issued by the same developer (hence, with the same MRSIGNER) can unseal it.

We note that Intel SGX has been deprecated in last-generation CPUs for desktops but will still be available for server-grade platforms~\cite{intelproducts}, which fits the confidential computing in the cloud paradigm.
Further, the size of the EPC will increase up to 1TB on multi-socket systems.

\subsection{Cloning SGX Enclaves}

As mentioned in our main paper~\cite{clonebuster_acsac}, cloning an application (irrespective of whether it resides within an enclave) may or may not include its runtime memory. ``Live'' cloning consists of creating a copy of a running process, that includes also the runtime memory of the original process. In contrast, a ``non-live'' cloning operation creates a clone by only copying the code and the persistent state.

We note that Intel SGX limits live cloning of enclaves ``by design''. In particular, EPC encrypted memory and hardware-managed EPCM prevent live cloning of enclaves: in a nutshell, an encrypted memory page assigned to a given enclave, cannot be copied and assigned to another one.

With respect to non-live cloning, we note that the sealing functionality used to persist state information to disk prevents cross-platform cloning. In particular, cryptographic keys that Intel SGX uses for sealing enclave data, depend on the host where the enclave is running. Therefore, state sealed by an enclave on a given host cannot be unsealed on a different host.

Nevertheless, Intel SGX does not prevent non-live cloning of an enclave on the same platform, nor does it provide a mechanism to distinguish two such clones. In particular, the number of enclaves that can be set up on a given host and executed at the same time---regardless of the loaded binary---is only limited by system resources. Thus, little prevents an adversary, that controls the OS on a given host, to launch a number of enclaves with the same binary. In case one of those enclaves seals data to disk, all other enclaves with the same binary have access to that data---since sealing keys on a given host only depend on the enclave identity. As a result, if one enclave is attested and provisioned with a secret, all clones will have access to the same secret. Intel acknowledges that there is no mechanism to distinguish enclaves loaded with the same binary on the same platform, since they all share the same identities (i.e., MRSIGNER and MRENCLAVE).\footnote{\url{https://intel.ly/3uprwdh}}

\subsection{\our{}}

While several solutions to thwart cloning attacks exist, they are either impractical—requiring monotonic counters, which are not available in the cloud, or relying on TTPs, which are hard to find in practice. We addressed this gap in our previous ACSAC'23 paper \cite{clonebuster_acsac}, where we present \our{}, the first practical cloning protection for Intel SGX that works directly within current off-the-shelf cloud deployments. \our{} relies on a cache covert channel to signal its presence to (and detect the presence of) clones without requiring special hardware. 
If the enclave instance is truly unique, it will see no response on the monitored channel. On the other hand, if multiple instances are running, each instance will observe a measurable response in the form of a contention pattern.

\our{} undergoes two phases of operation: a preparation phase and a monitoring phase. The preparation phase defines and sets up the ''channel'', i.e., selects a specific group of cache sets so that enclaves with the same (resp. different) binary will use the same (resp. a different) channel. Once the channel has been defined, \our{} builds the eviction sets required to communicate over such (cache-based) channels. During the monitoring phase, \our{} fills the cache sets of its channel with its own data and continuously measures the time to access such data to detect if it is still cached (cache hit) or has been evicted (cache miss). Clones will use the same channel (i.e., the same group of cache sets), removing each other’s data. The resulting sequence of cache hits and misses is then fed to a classifier whose role is to distinguish whether clones are running on the same host based on the input sequence.

Using a simple threshold-based classifier, \our{} achieves an F1 score up to 0.99 even in the worst-case scenario where an application with a high cache load is running in the background. At the same time, \our{} only incurs a marginal impact on the application performance and adds roughly 800 lines of code\footnote{\url{https://github.com/nec-research/CloneBuster}} to the Trusted Computing Base.

\section{Cloning Attacks on SGX Enclaves}
\label{sec:analysis}

As discussed in our main paper~\cite{clonebuster_acsac}, cloning has received comparatively little attention as an attack vector against SGX enclaves, especially when contrasted with rollback attacks. To address this gap, we investigated whether various SGX-based applications are susceptible to cloning attacks, highlighting a previously unnoticed risk. A more concise version of these results appeared in our ACSAC'23 paper~\cite{clonebuster_acsac}; here, we present the methodology and results in greater detail.

\subsection{Methodology}
To assess the impact of cloning attacks on SGX applications, we analyzed the security of 72 SGX-based applications against rollback and cloning attacks. Selected applications were taken from curated lists of SGX papers and applications~\cite{awesome_sgx,sgx_papers}---the most extensive collections of proposals leveraging SGX to increase application security. We excluded projects that provide libraries, frameworks, and attacks in our analysis. A cloning attack requires a concrete enclave instantiation; functions are not subject to cloning. Hence, libraries and similar are not subject to cloning attacks by default. Furthermore, we did not analyze projects without documentation. We divide the remaining ones based on the type of application (machine learning, blockchain, encrypted databases, etc.), following the structure from~\cite{awesome_sgx}. 

Our analysis is based on an individual investigation of the selected list of projects.
Here, we start by analyzing the description provided in the (white) paper where the proposal was introduced for each project.
To assess a proposal's susceptibility to rollback attacks, we examine whether it seals its state.

Note that, in our analysis, we consider a given application to be susceptible to rollback attacks or cloning attacks, if and only if, such attacks result in a meaningful exploitation of the state of the enclave. In particular, we do not consider DoS attacks to be within the scope of a meaningful exploitation.

\subsection{Results}

Our results are summarized in Table~\ref{tab:extended_analysis}.
For each application, we report whether the code is available, whether they are vulnerable to rollback attacks, and whether they are vulnerable to cloning attacks.
In case the application is not vulnerable to a specific attack, we report the countermeasure (MC is monotonic counters, and TTP is a trusted third party). We use N/A in case the attack is not applicable.
We find that all vulnerable applications can be attacked following one out of three schemes.
In case of applications vulnerable to cloning attacks, we categorize the attack type (A, B, C) and provide more details in Section~\ref{sec:attacks}. Proof of Luck ($\star$) is not vulnerable to cloning because it books all MCs on the platform at startup; as a result, no clone can be started, but this also means that MCs are no longer available for other applications on the same host.

Based on our findings, we draw the following observations:
\begin{itemize}
	\item Out of the 72 proposals, 14 applications (i.e., roughly 20\%) are vulnerable to forking attacks based on cloning.
	\item 11 of the vulnerable 14 applications do not account for any protection mechanism against forking attacks.
	\item 3 of the 14 vulnerable applications prevent rollback attacks with a monotonic counter; yet, they are vulnerable to forking attacks based on cloning.
	\item 7 of the 14 vulnerable applications do not seal state and, therefore, are not vulnerable to rollback attacks per design; however, those applications are vulnerable to cloning.
	\item Out of the 72 proposals, 11 use a TTP to prevent forking attacks. Among these 11 proposals, 9 rely on a decentralized ledger to prevent forking (8 of those are blockchain applications). Finally, 2 applications dismiss rollback attacks by claiming that these attacks can be mitigated by ROTE~\cite{ROTE2017}.
    \item 64.3\% of all database applications we analyzed are vulnerable to cloning attacks. Database applications maintain a large state, and although rollback is typically not an exploitable attack vector due to a monotonic counter, cloning here allows the attacker to present two users with different views of the same database. 
    \item 55 out of the 148 projects in the considered categories do not contain proper design documentation at the time of writing. We point out that extensive design documentation is required to ensure understanding the underlying design is feasible within an appropriate amount of time. 
    \item 58.3\% of the proposals provide no complete open-source implementation. 18 projects among the analyzed projects from both repositories provide no open-source implementation, and 24 projects provide an incomplete implementation for the design.
    
\end{itemize}

Our observations indeed confirm that cloning vulnerabilities are a real threat in SGX-based applications.

\begin{table*}[!htbp]
	\centering
	\footnotesize
    \noindent\scalebox{0.95}{
		\begin{tabular}{|l|c|l|l|l|l|c|l|l|}
				
			\cline{1-4} \cline{6-9}
			\multirow{2}{*}{\textbf{Project}} & \textbf{Source code} &  \multicolumn{2}{c|}{{\textbf{Vulnerable to}}} & & \multirow{2}{*}{\textbf{Project}} & \textbf{Source code} &  \multicolumn{2}{c|}{{\textbf{Vulnerable to}}} \\ \cline{3-4} \cline{8-9}
			& \textbf{available} &  \textbf{Rollback} & \textbf{Cloning} & & & \textbf{available} & \textbf{Rollback} & \textbf{Cloning} \\
		
			\cline{1-4} \cline{6-9}
			\multicolumn{4}{|c|}{\textbf{Encrypted Databases and Key-value Stores}} 						 				 & &  \cellcolor{gainsboro}\textbf{X-Search} \cite{xsearch_paper,xsearch_code} $^{ap}$ & \cellcolor{gainsboro}Yes & \cellcolor{gainsboro}N/A & \cellcolor{gainsboro}Yes (C) \\\cline{1-4} \cline{6-9}
			\cellcolor{gainsboro}\textbf{Aria} \cite{aria} $^{p}$ & \cellcolor{gainsboro}No & \cellcolor{gainsboro}N/A & \cellcolor{gainsboro}Yes (A)    									 				 & & \multicolumn{4}{c|}{\textbf{Blockchains}} \\ \cline{6-9}
			\cellcolor{gainsboro}\textbf{Avocado} \cite{avocado_paper,avocado_code} $^{a}$ & \cellcolor{gainsboro}Yes & \cellcolor{gainsboro}N/A & \cellcolor{gainsboro}Yes (A)    			 				 & & \textbf{BITE} \cite{bite} $^{p}$ & No  & No (MC) & N/A \\
			\cellcolor{gainsboro}\textbf{Enclage} \cite{enclage} $^{p}$ & \cellcolor{gainsboro}No & \cellcolor{gainsboro}N/A & \cellcolor{gainsboro}Yes (A)    								 				 & & \textbf{BLOXY} \cite{bloxy} $^{p}$ & No  & N/A & N/A \\
			\cellcolor{gainsboro}\textbf{EnclaveCache} \cite{enclavecache} $^{p}$ & \cellcolor{gainsboro}No & \cellcolor{gainsboro}Yes & \cellcolor{gainsboro}Yes (B)    					 				 & & \textbf{Ekiden} \cite{ekiden_paper,ekiden_code} $^{a}$ & Yes  & No (TTP) & No (TTP) \\
			\textbf{EnclaveDB} \cite{EnclaveDB} $^{p}$ & No  & No (MC) & No (TTP)    						 				 & & \textbf{Hybrids on Steroids} \cite{troxy} $^{p}$ & No  & No (MC+TTP) & No (TTP) \\
			\textbf{HardIDX} \cite{hardidx} $^{p}$ & No  & Yes & N/A    									 				 & & \textbf{MobileCoin} \cite{mobilecoin_paper,mobilecoin_code} $^{a}$ & Yes  & No (TTP)& No (TTP) \\
			\cellcolor{gainsboro}\textbf{NeXUS} \cite{nexus_paper,nexus_code} $^{p}$ & \cellcolor{gainsboro}Yes & \cellcolor{gainsboro}No (MC) & \cellcolor{gainsboro}Yes (B)    			 				 & & \textbf{Oasis} \cite{oasis} $^{a}$ & Yes  & No (TTP)& No (TTP)\\
			\cellcolor{gainsboro}\textbf{ObliDB} \cite{oblidb_paper,oblidb_code} $^{p}$ & \cellcolor{gainsboro}Yes & \cellcolor{gainsboro}N/A & \cellcolor{gainsboro}Yes (A)    				 				 & & \textbf{Obscuro} \cite{obscuro_paper,obscuro_code} $^{a}$ & Yes  & N/A & N/A \\
			\textbf{PESOS} \cite{pesos} $^{p}$ & No  & N/A & N/A    										 				 & & \textbf{Phala Network} \cite{phala_paper,phala_code} $^{a}$ & Yes  & No (TTP) & No (TTP) \\
			\textbf{SeGShare} \cite{segshare} $^{p}$ & No  & No (MC) & N/A    								 				 & & \textbf{Private Chaincode} \cite{fabric_paper,fabric_code} $^{a}$ & Yes  & N/A & N/A \\
			\cellcolor{gainsboro}\textbf{ShieldStore} \cite{shieldstore_paper,shieldstore_code} $^{ap}$ & \cellcolor{gainsboro}Yes  & \cellcolor{gainsboro}No (MC) & \cellcolor{gainsboro}Yes (B)  			 & & \textbf{Private Data Objects} \cite{pdo_paper,pdo_code} $^{a}$ & Yes  & No (TTP) & No (TTP) \\
			\textbf{SPEICHER} \cite{speicher_paper,speicher_code} $^{a}$ & Yes  & No (MC) & No (TTP)    	 				 & & \textbf{Proof of Luck} \cite{pol_paper,pol_code} $^{a}$ & Yes  & N/A & No ($\star$) \\
			\cellcolor{gainsboro}\textbf{STANlite} \cite{stanlite_paper,stanlite_code} $^{ap}$ & \cellcolor{gainsboro}Yes  & \cellcolor{gainsboro}N/A & \cellcolor{gainsboro}Yes (A)    		 				 & & \textbf{Teechain} \cite{teechain_paper,teechain_code} $^{ap}$ & Yes  & No (MC) & N/A \\
			\cellcolor{gainsboro}\textbf{StealthDB} \cite{stealthdb_paper,stealthdb_code} $^{a}$ & \cellcolor{gainsboro}Yes  & \cellcolor{gainsboro}Yes & \cellcolor{gainsboro}Yes (B)    	 				 & & \textbf{Town Crier} \cite{towncrier_paper,towncrier_code} $^{a}$ & Yes  & No (TTP) & No (TTP) \\
			
			\cline{1-4}
			\multicolumn{4}{|c|}{\textbf{Applications}}														 & & \textbf{Troxy} \cite{troxy} $^{p}$ & No  & N/A & No (TTP) \\ \cline{1-4}
			\cellcolor{gainsboro}\textbf{BI-SGX} \cite{bisgx_code} $^{a}$ & \cellcolor{gainsboro}Yes  & \cellcolor{gainsboro}No (MC) & \cellcolor{gainsboro}Yes (B) 							 & & \textbf{Twilight} \cite{twilight_paper,twilight_code} $^{a}$ & Yes  & N/A & N/A \\ \cline{6-9}
            \cellcolor{gainsboro}\textbf{CACIC} \cite{cacic_paper,cacic_code} $^{a}$ & \cellcolor{gainsboro}Yes  & \cellcolor{gainsboro}Yes & \cellcolor{gainsboro}Yes (B) 					 & & \multicolumn{4}{c|}{\textbf{Machine Learning}} \\ \cline{6-9}
			\textbf{DEBE} \cite{debe_paper, debe_code} $^{a}$ & Yes  & N/A & N/A 						 	 & & \textbf{Confidential ML} \cite{confidential_ml} $^{a}$ & Yes  & N/A & N/A\\
			\textbf{HySec-Flow} \cite{hysec_paper,hysec_code} $^{a}$ & Yes  & N/A & N/A 					 & & \textbf{DP-GBDT} \cite{dp_gbdt} $^{a}$ & Yes  & No (MC) & N/A \\
			\cellcolor{gainsboro}\textbf{PrivaTube} \cite{privatube} $^{p}$ & \cellcolor{gainsboro}No  & \cellcolor{gainsboro}N/A & \cellcolor{gainsboro}Yes (C) 								 & & \textbf{Plinius} \cite{plinius_paper,plinius_code} $^{a}$ & Yes  & No (MC) & N/A \\
			\textbf{REX} \cite{rex_paper, rex_code} $^{a}$ & Yes  & N/A & N/A 								 & & \textbf{secureTF} \cite{securetf} $^{p}$ & No  & N/A & N/A \\
			\textbf{SGXDedup} \cite{dedup_paper,dedup_code} $^{a}$ & Yes  & N/A & N/A 						 & & \textbf{Secure XGBoost} \cite{xgboost_paper,xgboost_code} $^{a}$ & Yes & N/A & N/A \\
			\textbf{Signal CDS} \cite{signal_blog,signal_code} $^{a}$ & Yes  & N/A & N/A 					 & & \textbf{Slalom} \cite{slalom_paper,slalom_code} $^{ap}$ & Yes  & N/A & N/A \\
			\textbf{SkSES} \cite{skses_paper,skses_code} $^{a}$ & No & N/A & N/A 							 & & \textbf{SOTER} \cite{soter_paper,soter_code} $^{a}$ & Yes  & N/A & N/A \\ \cline{6-9}
			\textbf{SMac} \cite{smac_gen_paper,smac_gen_code} $^{a}$ & Yes  & N/A & N/A 					 & & \multicolumn{4}{c|}{\textbf{Network}} \\ \cline{6-9}
			\textbf{TresorSGX} \cite{tresorsgx_paper,tresorsgx_code} $^{a}$ & Yes  & N/A & N/A 				 & & \textbf{ConsenSGX} \cite{consensgx_paper,consensgx_code} $^{a}$ & No  & N/A & N/A \\
			
			\cline{1-4}
			\multicolumn{4}{|c|}{\textbf{Key + Password Management}} 										 & & \textbf{CYCLOSA} \cite{cyclosa} $^{p}$ & No  & N/A & N/A \\ \cline{1-4}
			\textbf{DelegaTEE} \cite{delegatee} $^{p}$ & No  & No (MC) & N/A 								 & & \textbf{ENDBOX} \cite{endbox} $^{ap}$ & No  & N/A & N/A \\
			\textbf{FeIDo} \cite{feido_paper,feido_code} $^{a}$ & Yes  & N/A & N/A 							 & & \textbf{LightBox} \cite{lightbox_paper,lightbox_code} $^{ap}$ & Yes  & N/A & N/A \\
			\textbf{Keys in Clouds} \cite{keyscloud_paper,keyscloud_code} $^{a}$ & Yes  & No (MC) & N/A		 & & \textbf{SENG} \cite{seng_paper,seng_code} $^{a}$ & Yes  & N/A & N/A \\
			\textbf{SafeKeeper} \cite{safekeeper_paper,safekeeper_code} $^{a}$ & Yes  & No (MC) & N/A 		 & & \textbf{SGX CBR} \cite{sgxcbr} $^{p}$ & No  & N/A & N/A \\
			\cellcolor{gainsboro}\textbf{SGX-KMS} \cite{sgxkms_paper,sgxkms_code} $^{a}$ & \cellcolor{gainsboro}Yes  & \cellcolor{gainsboro}Yes & \cellcolor{gainsboro}Yes (B) 				 & & \textbf{SGX-Tor} \cite{sgxtor_paper,sgxtor_code} $^{ap}$ & Yes  & Yes & N/A \\
			
			\cline{1-4}
			\multicolumn{4}{|c|}{\textbf{Private Search}} 													 & & \textbf{TEE V2V} \cite{v2v_paper,v2v_code} $^{a}$ & No  & N/A & N/A \\ \cline{1-4}
			\textbf{BISEN} \cite{bisen_paper,bisen_code} $^{a}$ & Yes & N/A & N/A 							 & & \textbf{MACSec} \cite{macsec} $^{a}$ & Yes  & No (MC) & N/A \\
			\textbf{DeSearch} \cite{desearch_paper,desearch_code} $^{a}$ & Yes  & N/A & N/A 				 & & \textbf{S-NFV} \cite{snfv} $^{p}$ & No  & N/A & N/A \\
			\textbf{Maiden} \cite{maiden_paper,maiden_code} $^{a}$ & Yes & N/A & N/A 						 & & \textbf{SafeBricks} \cite{safebricks_paper,safebricks_code} $^{ap}$ & Yes  & N/A & N/A \\
			\textbf{POSUP} \cite{posup_paper,posup_code} $^{a}$ & Yes & N/A & N/A 					         & & \textbf{SELIS-PubSub} \cite{pubsub_paper,pubsub_code} $^{p}$ & Yes  & N/A & N/A \\ \cline{6-9}
			\textbf{QShield} \cite{qshield_paper,qshield_code} $^{a}$ & Yes  & N/A & N/A 					 & & \multicolumn{4}{c|}{\textbf{Data Analytics}} \\ \cline{6-9}
			\textbf{Snoopy} \cite{snoopy_paper,snoopy_code} $^{a}$ & Yes  & No (MC) & N/A 					 & & \textbf{Opaque} \cite{opaque_paper,opaque_code} $^{a}$ & Yes  & No (TTP) & No (TTP) \\ 					
			\cline{1-4} \cline{6-9}
		\end{tabular}
	}
	\caption{
		Summary of our analysis of SGX applications. We analyzed SGX applications listed in~\cite{sgx_papers} (superscript $p$ next to the citation) and listed in~\cite{awesome_sgx} (superscript $a$ next to the citation). Applications that are vulnerable to cloning attacks are highlighted in gray.
    }
	\label{tab:extended_analysis}
\end{table*}

\section{The Three Classes of Cloning Attacks}
\label{sec:attacks}

We observe that cloning-based attacks can be grouped into three broad categories, namely, A, B, and C. Category A mostly consists of in-memory key-value stores (KVS); by cloning the application, the adversary can split the inputs from different clients across multiple KVS instances so that clients have different ``views'' of the store (e.g., updates made by one client to a specific key are not seen by another client). Applications in category B seal state to have it available across restarts; by cloning these applications, the adversary can obtain multiple valid states that can be fed to the enclave when it restarts. Category C mostly consists of applications that leverage an SGX enclave as a proxy to guarantee the unlinkability or privacy of client requests; by cloning the application, the adversary can partition the set of clients, thereby reducing the anonymity set for each client.
We now describe in detail how to mount cloning attacks on these applications. To ease the presentation, we describe each of the attack categories and, for each category, we show how the attack can be mounted against an exemplary application (chosen from the 14 vulnerable applications).
	
\subsection{\catA ~-- Forking In-memory Key-value Stores}
We start by describing cloning attacks on in-memory key-value stores (KVS), dubbed \catA.
A KVS exposes \texttt{PUT} and \texttt{GET} interfaces to clients. At any time, there is at most one value per key; further, when a \texttt{GET} request is issued for a given key $k$, the latest value that was written to the database for that key is returned.
A forking attack against a KVS may break these security guarantees.
	
In-memory KVSs that use Intel SGX, e.g., Aria \cite{aria}, Enclage \cite{enclage}, STANLite \cite{stanlite_paper}, ObliDB \cite{oblidb_paper}, and Avocado \cite{avocado_paper}, are designed to store data larger than EPC memory (limited to 128 MB). Hence, they seal data to persistent memory but keep meta-data in runtime memory to ensure integrity and rollback protection. Since meta-data is not sealed to persistent storage, it is lost if the enclave terminates.
	
\subsubsection{Attack overview}
	
\begin{figure} [!t]
	\centering 
	\includegraphics[width=0.41\textwidth]{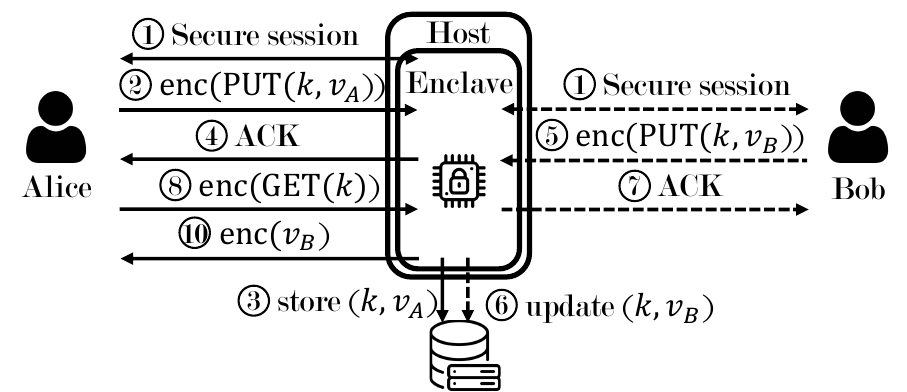}
	\caption{Overview of an SGX-backed in-memory key-value store if it operates in a benign setting.}
	\label{fig:fim_1}
\end{figure}

In benign settings, assume a server running an enclave-backed KVS that two clients, $A$ and $B$, can access as depicted in Figure \ref{fig:fim_1}.
First, both clients attest the enclave and establish a session key to encrypt their messages.
All subsequent messages are encrypted using the session key.
In a benign setting, $A$ sends a \texttt{PUT} request to post the key-value (KV) pair $(k, v_A)$ to the storage.
The enclave recognizes that the key $k$ does not exist, creates a new entry with the pair $(k, v_A)$, and returns an $ACK$ message.
Afterward, $B$ sends a \texttt{PUT} request to post the KV pair $(k, v_B)$.
At this time, the enclave recognizes that the key $k$ exists and updates the value to $v_B$ before returning an $ACK$ message.
If $A$ later requests the value associated with $k$ from the KVS, it receives the value $v_B$.
	
\begin{figure} [!t]
	\centering
	\includegraphics[width=0.42\textwidth]{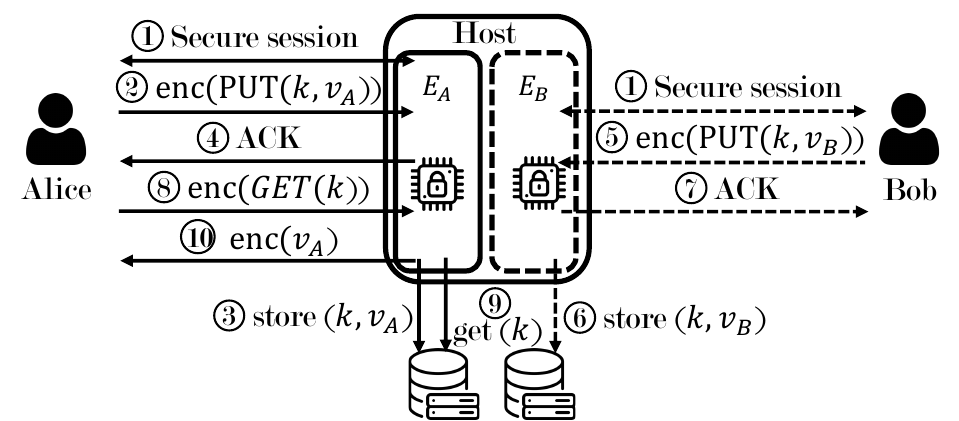}
	\caption{Overview of a generic \catA ~attack on an SGX-backed in-memory key-value store.}
	\label{fig:fim_2}
\end{figure}

\pagebreak
In an adversarial setting (cf. Figure \ref{fig:fim_2}), the adversary can provide two KVS instances to $A$ and $B$.
Even if clients attest the enclave where the KVS instance is running, they cannot tell whether they are communicating with the same instance or not.
	
The adversary launches two enclave instances, $E_A$ and $E_B$, and connects each client to one instance.
Each client attests the connected enclave.
Assume the same sequence of requests as in the previous setting.
First, $A$ sends a \texttt{PUT} request to post the KV pair $(k, v_A)$ to the storage.
$E_A$ recognizes that $k$ does not exist in the associated storage, creates a new entry with the pair $(k, v_A)$, and returns an $ACK$ message.
Afterward, client $B$ sends a \texttt{PUT} request to post the KV pair $(k, v_B)$.
$E_B$ does not find an entry for $k$ in its copy of the storage and creates a new entry for the pair $(k, v_B)$.
Both clients receive an $ACK$ reporting the correct execution of their request.
However, if client $A$ later requests the value associated with $k$, $E_A$ returns $v_A$ which is the latest value it has seen---this is however different from the newest value in the system.
	
We note that the above issue can be fixed if clients are mutually trusted and each client shares its view of the KVS with the others~\cite{brandenburger17dsn}. Essentially, the set of clients acts as a distributed TTP. Unfortunately, assuming a mutually trusted set of clients limits the scenarios where the KVS can be used.
	
\subsubsection{Concrete example: \catA ~Attack against Aria}
	
As an example, we now show how to mount a \catA ~attack against \textbf{Aria} \cite{aria}.
Aria provides an in-memory KVS in the cloud.
Each entry is protected against rollback attacks by means of  MC.
For confidentiality, the entries are encrypted with AES (in CTR mode), where the counter value is set to be the current MC value of the entry.
The enclave generates a pseudo-random key at initialization and uses the same key for encrypting all data.
Additionally, each entry contains a MAC over the encrypted data for integrity protection.
The integrity of the MCs is guaranteed by a Merkle tree structure over all MCs.
The enclave exclusively stores the Merkle root in its runtime memory.
Additionally, it stores all recently used MCs in its local cache.
The cached counters can be used to decrypt entries directly without verifying the Merkle root, thus reducing latency.
	
\begin{figure}
	\centering
	\includegraphics[width=0.45\textwidth]{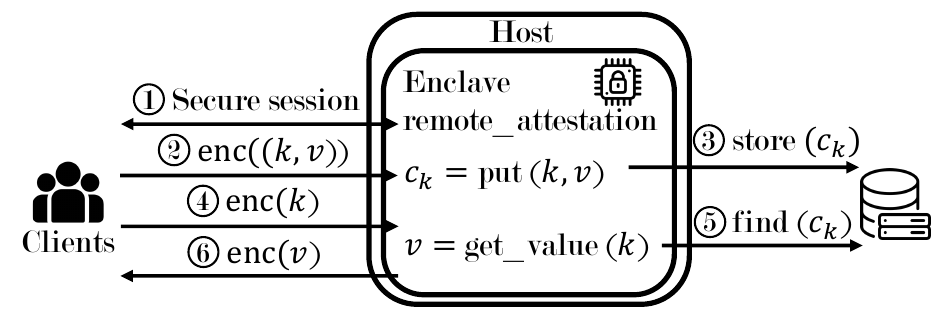}
	\caption{Overview of the main functions exposed by the Aria enclave and its interactions with clients.}
	\label{fig:aria_1}
\end{figure}

As depicted in Figure \ref{fig:aria_1}, the client first attests the enclave and establishes a secure session key.
The client sends a \texttt{PUT} request for $(k, v)$, encrypted with the session key.
The enclave decrypts the message and checks if $k$ exists in storage.
If so, it updates the corresponding counter and encrypted KV pair.
Otherwise, it assigns the key a free counter and stores it in the database.
Later, the client can access $v$ by sending an encrypted \texttt{GET} request for $k$.
The enclave verifies the counter integrity and decrypts entries until it finds the requested KV pair.
Finally, it returns $v$ through the secure channel.
	
\begin{figure}
	\centering
	\includegraphics[width=0.39\textwidth]{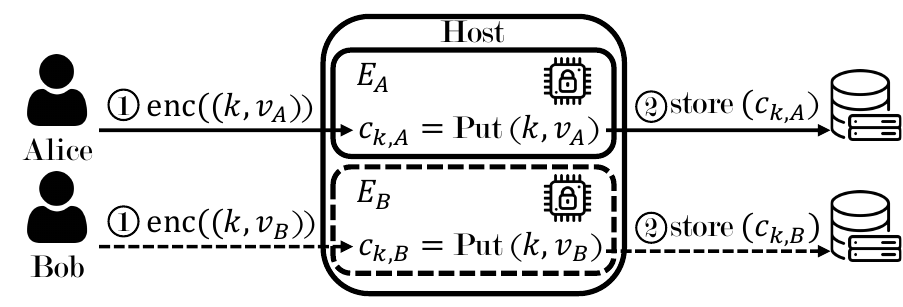}
	\caption{Overview of a cloning attack against Aria enclaves.}
	\label{fig:aria_2}
	\vspace{-1em}
\end{figure}

Assume now a malicious host and two clients, $A$ and $B$ who share access to the same KVS, e.g. for customer records.
As shown in Figure \ref{fig:aria_2}, one can mount \catA ~attacks on Aria as follows:
\begin{itemize}
	\item The adversary starts two Aria enclave instances, $E_A$ and $E_B$.
	\item The adversary connects $A$ to $E_A$, and $B$ to $E_B$.
	\item The clients attest the enclaves and establish secure communication sessions.
	\item The clients send encrypted \texttt{PUT} requests for $(k, v_A)$ and $(k, v_B)$ to $E_A$ and $E_B$, respectively.
	\item $E_A$ and $E_B$ decrypt the requests and create/update the corresponding encrypted entries in their storage instances. $A$ and $B$ cannot determine if they are communicating with the same instance.
\end{itemize}
	
Hence, \catA ~violates the consistency of Aria by cloning the enclave.
The adversary is not limited by the number of enclaves and can run arbitrarily many instances.
	
\subsection{\catB ~-- Forking persistent Key-value Stores}
	
We now describe cloning attacks on persistent KVSs, dubbed \catB.
Persistent KVSs that are susceptible to cloning attacks are EnclaveCache \cite{enclavecache}, NeXUS \cite{nexus_paper}, StealthDB \cite{stealthdb_paper}, ShieldStore \cite{shieldstore_paper}, SGX-KMS \cite{sgxkms_paper}, and CACIC \cite{cacic_paper}.
In contrast to in-memory KVSs, persistent KVSs use sealing to make meta-data available across reboots.
	
\subsubsection{Attack overview}

\begin{figure}[t]
	\centering
	\includegraphics[width=0.26\textwidth]{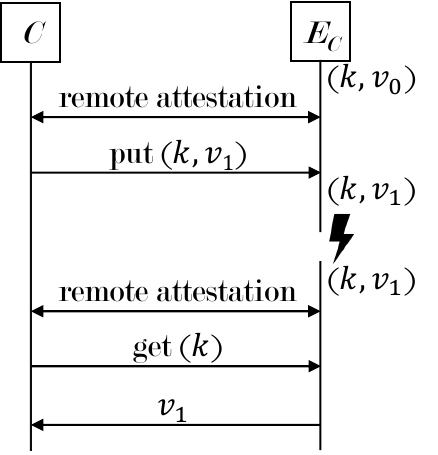}
	\caption{Overview of the interaction of a persistent key-value store with a client in a benign setting.}
	\label{fig:forkvs_1}
\end{figure}

Assume a server running an enclave-backed KVS that stores a KV pair $(k, v_0)$ when a client, $C$, connects to the system.
First, $C$ attests the enclave $E_C$ and establishes a session key.
All subsequent messages are encrypted using the session key.
In a benign setting (cf. Figure~\ref{fig:forkvs_1}), $C$ sends a \texttt{PUT} request to update the value associated with $k$ to $v_1$.
$E_C$ updates the KV pair in its storage.
Afterward, it creates a snapshot, sealing the meta-data with the incremented MC value $ctr_i + 1$ and incrementing the MC value $ctr_{i+1} \leftarrow ctr_{i} + 1$.
Later, $E$ crashes and needs to restart.
It successfully verifies the MC value in the sealed data and restores the KVS.
$C$ must attest the restarted enclave instance and establish new session keys.
When $C$ requests the value associated with $k$, the KVS correctly returns the latest value, $v_1$.
	
\begin{figure}[t]
	\centering
	\includegraphics[width=0.35\textwidth]{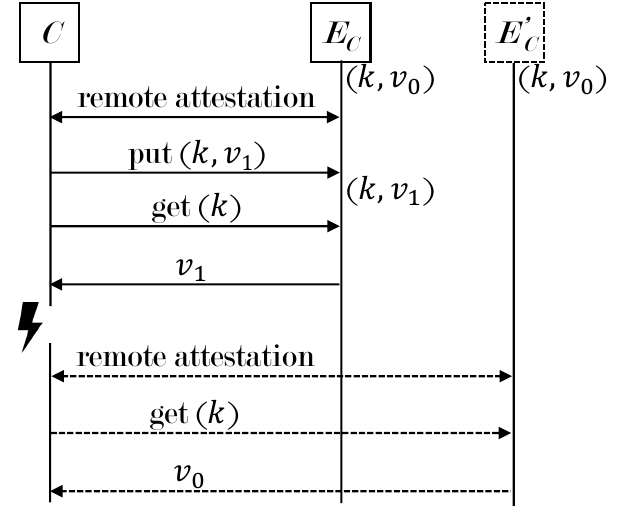}
	\caption{Overview of a generic \catB ~attack against persistent key-value stores.}
	\label{fig:forkvs_2}
\end{figure}

In an adversarial setting (cf. Figure \ref{fig:forkvs_2}), the adversary can provide two different views of the same KVS instance to $C$.
The adversary launches two enclave instances, $E_C$ and $E'_C$.
Both instances have the same initial state storing the KV pair $(k, v_0)$.
First, the adversary connects $C$ to enclave $E_C$, and the value is updated to $v_1$.
If the client requests the value associated to $k$, the enclave correctly returns $v_1$.
Afterward, the adversary connects $C$ to the second instance, $E'_C$.
$C$ assumes the enclave has crashed and successfully attests $E'_C$.
However, when requesting the value associated with $k$, $E'_C$ returns $v_0$, the latest state it stores.
Consequently, the same key is associated with different values in different enclave instances.
The same attack holds if multiple clients use the KVS instead of $C$ connecting to the KVS in different sessions.
	
\vfill\null
Cloning attacks on in-memory KVSs are limited to providing two instances of a KVS.
They do not share entries unless the same data is provided to both instances in different sessions.
In contrast, \catB ~is more powerful: two instances of the KVS share common data that has been sealed by the first instance before the second instance starts.
Therefore, \catB ~can have the same effect as rollback attacks, even if classical rollback attacks are not possible.
	
\subsubsection{Concrete example: \catB ~Attack against BI-SGX}

As an example, we show how to successfully mount a forking attack based on cloning against \textbf{BI-SGX}~\cite{bisgx_code}, as previously presented in our ACSAC'23 paper~\cite{clonebuster_acsac}. 
BI-SGX provides secure computation over private data in the cloud by leveraging SGX. 

\begin{figure}
	\centering
	\includegraphics[width=0.5\textwidth]{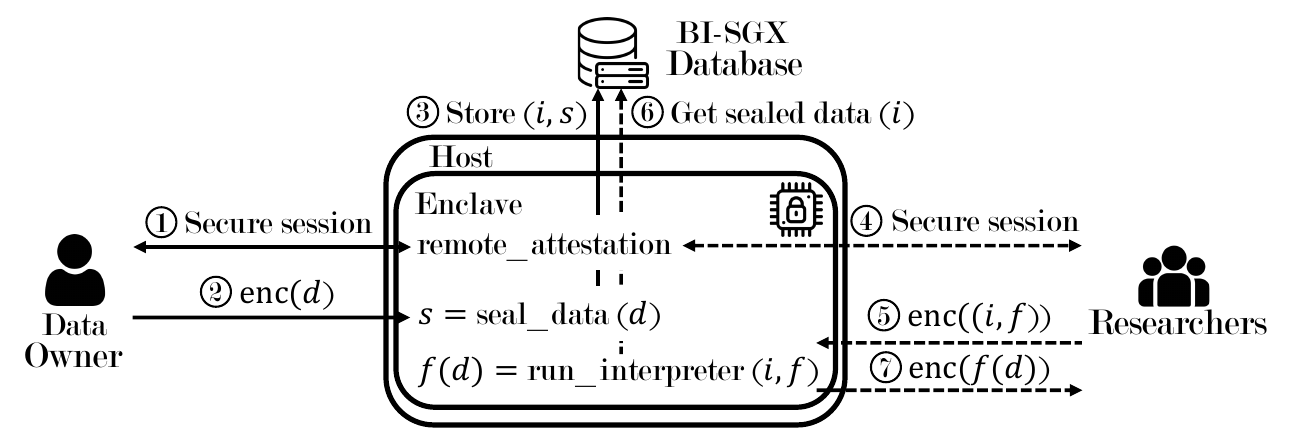}
	\caption{Overview of the main functions exposed by the BI-SGX enclave and its interactions with Data Owners and Researchers.}
	\label{fig:bisgx}
\end{figure}

As shown in Figure~\ref{fig:bisgx}, a \textit{data-owner} sends to the BI-SGX enclave data $d$ encrypted as ciphertext $c_O$;
the encryption key is agreed between the enclave and the data owner via remote attestation.
The BI-SGX enclave decrypts the plaintext, seals it, and sends the sealed data (denoted as $s$) to an external database. 
The database stores $s$ along with an index $i$ as a tuple $(i, s)$.
Later on, a \textit{researcher} can send requests to the enclave; requests include the index that is used to retrieve data from the database and a description of a function $f$ to be computed over the data. 
More precisely, a request includes a tuple $(i,f)$ so that if $(i, s)$ exists in the database, the enclave unseals $s$ and returns $f(d)$.

As stated by Jangid et al.~\cite{jangid2021towards}, queries issued by researchers (and containing different indexes) should retrieve and process different data or, the other way around, queries containing the same index should process the same data.
BI-SGX cannot guarantee such a property, i.e., an attacker can feed the enclave with different data even if researchers submit requests with the same index.
The index used for data retrieval is not included in the sealed data but added by the database when it receives the encrypted data for storage. 
Upon request issued by the BI-SGX enclave to retrieve data item with index $i$, a malicious OS could return any sealed data item; 
the enclave has no means to tell if the sealed data returned by the OS is the right one.

The vulnerability was reported to the developers of BI-SGX by Jangid et al.~\cite{jangid2021towards}. 
The latter also proposed to use monotonic counters to mitigate this attack. 
The idea is to seal the index of the data along with the data itself. 
Hence, when the BI-SGX enclave requests sealed data with index $i$ and obtains a ciphertext $Enc(d,j)$, it only accepts $d$ as valid if $i=j$. 
Further, the use of MCs as indexes ensure that not two data items can be stored with the same index.

\pagebreak
Assume now a malicious server.
Even if the fix is implemented as suggested by \cite{jangid2021towards} and in \textit{inc-then-store} mode, one can mount a \catB ~attack against BI-SGX as follows:

\begin{figure}
	\centering
	\includegraphics[width=0.48\textwidth]{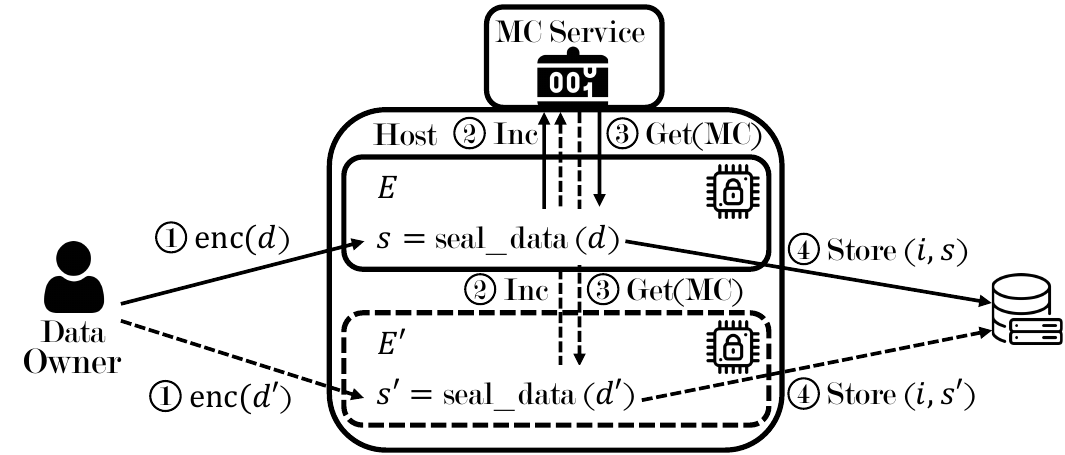}
	\caption{Overview of a cloning attack against the fixed BI-SGX version.}
	\label{fig_bisgx_att}
\end{figure}

\begin{itemize}
	\item The adversary starts two BI-SGX enclave instances (cf. Figure~\ref{fig_bisgx_att}).
	\item The adversary feeds one data item $d$ to enclave $E$ and another data item $d'$ to enclave $E'$. The current value of the counter is $MC$.
	\item The adversary stops the instance that first executes \texttt{Increment(MC)} until the other one has also executed it. The counter at this state is equal to $MC+2$. In practice an attacker could use a framework such as SGX-Step \cite{sgxstep17}.
	\item The adversary allows both instances to proceed. They execute \texttt{Read(MC)} and get exactly the same value of the counter ($MC+2$).
	\item Instance $E$ seals $(d,MC+2)$ while instance $E'$ seals $(d',MC+2)$. Both ciphertexts are sent to the database. Both ciphertexts are valid for a query from a \textit{researcher} to process data stored at index $MC+2$, as the BI-SGX enclave only checks if $MC$ in the sealed blob is equal to the index value in the \textit{researcher} request.
\end{itemize}

Hence, the attack violates the consistency of BI-SGX by cloning the enclave.

\subsection{\catC ~-- Breaking Unlinkability Guarantees}

We now describe cloning attacks on SGX proxies, dubbed \catC.
Applications affected by \catC, i.e., X-Search \cite{xsearch_paper} and PrivaTube \cite{privatube}, provide unlinkability by leveraging an SGX-backed proxy.
The proxy receives encrypted requests and obfuscates them, e.g., by adding fake requests, to ensure that an adversary accessing the service cannot link the plaintext requests to individual clients.
	
\subsubsection{Attack overview}
	
By cloning the enclave, an adversary can break the unlinkability guarantee and link a request to a specific user or at least reduce the anonymity set.
We now describe the generic cloning attack for breaking unlinkability guarantees of SGX-backed proxies, considering an honest setting first.
	
\begin{figure}
	\centering
	\includegraphics[width=0.42\textwidth]{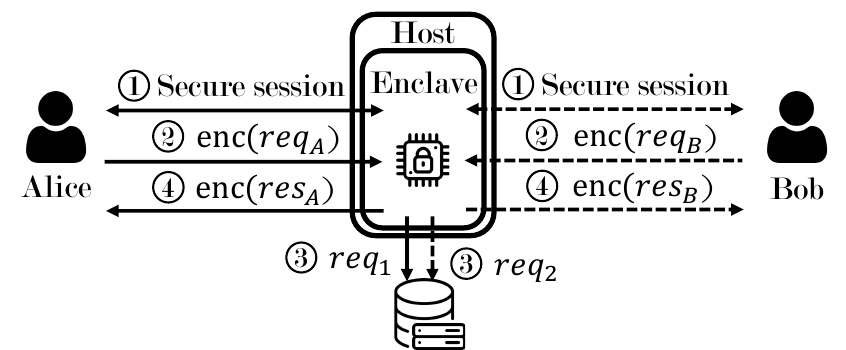}
	\caption{Overview of an SGX-backed proxy in a benign setting.}
	\label{fig:bug_1}
\end{figure}

Assume a server running an enclave-backed proxy that receives requests from two clients, $A$ and $B$, as depicted in Figure \ref{fig:bug_1}.
First, both clients attest the enclave and establish session keys.
In a benign setting, clients send requests $req_A$ and $req_B$, encrypted with the session key.
The enclave decrypts the requests and forwards two (decrypted) requests, $req_1$ and $req_2$, without knowing the identity of the issuer.
Afterward, the proxy maps the responses to the client requests, encrypts and forwards them to the corresponding client.
The server cannot distinguish if $A$ send $req_1$ or $req_2$.
The anonymity set increases with the number of clients simultaneously connected to the enclave.
	
\begin{figure}
	\centering
	\includegraphics[width=0.42\textwidth]{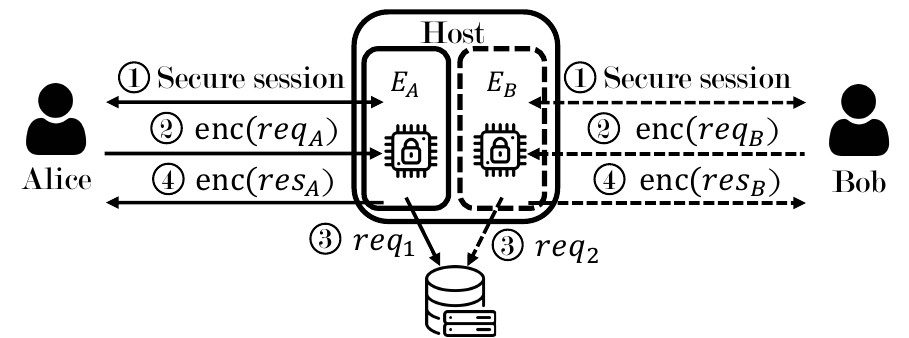}
	\caption{Overview of a generic \catC ~attack against an SGX-backed proxy.}
	\label{fig:bug_2}
\end{figure}

In an adversarial setting (cf. Figure \ref{fig:bug_2}), the adversary can recover the assignment and break the unlinkability guarantee.
The adversary starts two proxy enclaves, $E_A$ and $E_B$, and connects clients $A$ and $B$ to one instance each.
The clients attest the connected enclave and send the encrypted requests.
The adversary observes which enclave forwards the request to the server, e.g., $E_A$ sends the request $req_1$.
The adversary connected $A$ to $E_A$, thus can infer that $A$ sent $req_1$.
Linking the decrypted requests to the clients, the \catC ~attack breaks the unlinkability guarantee.
	
\subsubsection{Concrete example: \catC ~Attack against PrivaTube proxies}
	
As an example, we now show how to mount a \catC ~attack against \textbf{PrivaTube} \cite{privatube}.
PrivaTube is a distributed Video on Demand system leveraging fake requests and SGX enclaves to ensure the unlinkability of requests to individual users.
Requests for video segments can be served by video servers and assisting platforms.
Assisting platforms are other users that requested a specific video segment in the past and can provide other users with this segment.
Each peer in the system hosts an enclave, an HTTP proxy, to break the link between clients and requests.

\begin{figure}[t]
	\centering
	\includegraphics[width=0.50\textwidth]{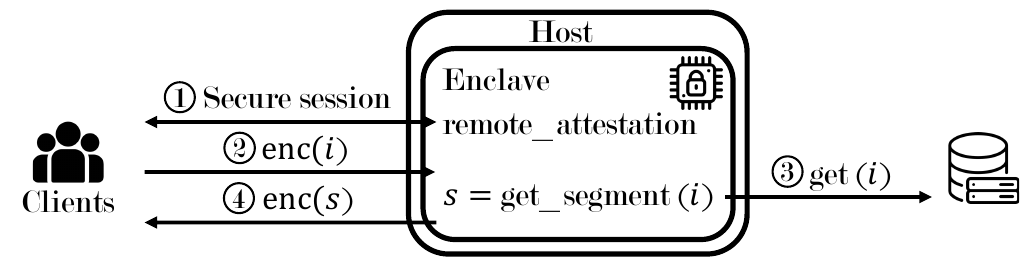}
	\caption{Overview of the functions exposed by a PrivaTube proxy and its interaction with clients.}
	\label{fig:privatube_1}
\end{figure}

As shown in Figure \ref{fig:privatube_1}, a client attests the proxy enclave and sends an encrypted request for a video segment with the ID $i$.
The enclave decrypts the segment ID and requests the video segment from the peer's video database.
It encrypts the received segment $\texttt{s}$ and sends it to the client.
PrivaTube assumes that each video server serves multiple requests simultaneously, thus preventing the precise assignment of users to requested video segments.
	
\begin{figure}[t]
	\centering
	\includegraphics[width=0.48\textwidth]{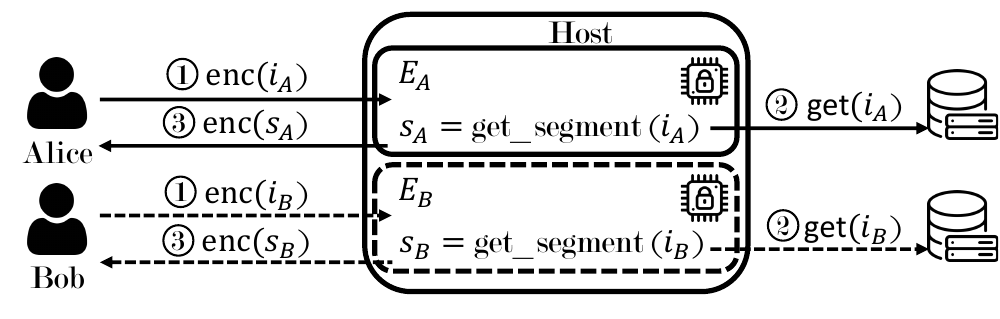}
	\caption{Overview of a cloning attack against a PrivaTube proxy.}
	\label{fig:privatube_2}
\end{figure}

Assume now a malicious video server and two users, $A$ and $B$.
As shown in Figure \ref{fig:privatube_2}, one can mount a \catC ~attack on PrivaTube proxies as follows:

\begin{itemize}
	\item The adversary starts two proxy enclave instances, $E_A$ and $E_B$.
	\item The adversary connects $A$ to $E_A$, and $B$ to $E_B$.
	\item The clients attest the enclaves and establish secure communication sessions.
	\item The clients send encrypted requests for $i_A$ and $i_B$ to $E_A$ and $E_B$, respectively.
	\item The adversary observes the decrypted requests for $i_A$ and $i_B$ to the database, issued by $E_A$ and $E_B$, respectively.
	\item Knowing $A$ is connected to $E_A$ and $B$ is connected to $E_B$, the adversary can recover that $A$ requested the video segment $i_A$ and $B$ requested $i_B$.
	Both requests are served correctly.
	Further, $A$ and $B$ cannot determine that they are connected to different proxies.
\end{itemize}
	
The adversary is not limited by the amount of enclaves it can execute at the same time.
For every client requesting the video server, the adversary can start a new enclave, precisely recovering the assignment of requested video segments to clients.
Here, the unlinkability guarantee is broken.
	
Notice that \catC{} attacks can be mounted against other proposals (not present in the lists we have analyzed) that use Intel SGX to cloak client requests. For example, Prochlo~\cite{prochlo} and subsequent work (e.g.,~\cite{orshuffle}) provide real-world privacy-preserving analytic frameworks to collect anonymous statistics. They use Intel SGX to shuffle client inputs so to break the link between a data item and the identity of the client where data originates. In a nutshell, an external observer cannot tell the client that sent a given input, among a set of $k$ clients. We note that, by cloning the shuffler enclave, an adversary can split the set of clients in disjoint subsets, so that two clients of two different subsets send their inputs to two different instances of the shuffler enclave. As a result, each client input is shuffled with less than $k$ data items and matching an input to its client becomes an easier task.

\section{Conclusion}
In this work, we addressed the problem of forking attacks against Intel SGX by cloning the victim enclave. We analyzed 72 SGX-based applications and found that roughly 20\% are vulnerable to such attacks, showing the detrimental impact of cloning attacks on SGX applications. We showed that even if applications rely on monotonic counters to prevent rollback attacks, they might still be susceptible to cloning attacks. 
We further observed that cloning-based attacks can be grouped into three broad categories. The first category targets the consistency of in-memory key-value stores, the second category targets persistent key-value stores, and the third category breaks the unlinkability guarantees of SGX proxies. Our results show that database applications seem to be particularly susceptible to cloning attacks as we found 64\% of all database applications vulnerable.

\section*{Acknowledgment}
This work is partly funded by the Deutsche Forschungsgemeinschaft
(DFG, German Research Foundation) under Germany’s
Excellence Strategy - EXC 2092 CASA - 390781972,
and the
European Union’s Horizon 2020 research and innovation programme (SPATIAL, Grant Agreement No. 101021808; REWIRE, Grant Agreement No. 101070627).
Views
and opinions expressed are however those of the authors
only and do not necessarily reflect those of the European
Union. Neither the European Union nor the granting
authority can be held responsible for them.

\bibliographystyle{IEEEtranS}
\bibliography{references}

@inproceedings{clonebuster_acsac, 
author = {Briongos, Samira and Karame, Ghassan and Soriente, Claudio and Wilde, Annika}, 
title = {No Forking Way: Detecting Cloning Attacks on Intel SGX Applications}, 
year = {2023}, isbn = {9798400708862}, 
publisher = {Association for Computing Machinery}, 
address = {New York, NY, USA}, 
url = {https://doi.org/10.1145/3627106.3627187}, 
doi = {10.1145/3627106.3627187},  
booktitle = {Proceedings of the 39th Annual Computer Security Applications Conference}, 
pages = {744–758}, 
numpages = {15}, 
location = {, Austin, TX, USA, }, 
series = {ACSAC '23} }

@inproceedings{DBLP:conf/ndss/Kaptchuk0M19,
  author       = {Gabriel Kaptchuk and
                  Matthew Green and
                  Ian Miers},
  title        = {Giving State to the Stateless: Augmenting Trustworthy Computation
                  with Ledgers},
  booktitle    = {Network and Distributed System Security Symposium, ({NDSS})},
  pages        = {1--15},
  year         = {2019},
}

@inproceedings{DBLP:conf/ccs/DijkRSD07,
  author       = {Marten van Dijk and
                  Jonathan Rhodes and
                  Luis F. G. Sarmenta and
                  Srinivas Devadas},
  title        = {Offline untrusted storage with immediate detection of forking and
                  replay attacks},
  booktitle    = {{ACM} Workshop on Scalable Trusted Computing ({STC})},
  pages        = {41--48},
  year         = {2007},
}

@inproceedings{NARRATOR2022,
  author    = {Jianyu Niu and Wei Peng and Xiaokuan Zhang and Yinqian Zhang},
  title     = {{NARRATOR:} Secure and Practical State Continuity for Trusted Execution
               in the Cloud},
  booktitle = {Proceedings of the 2022 {ACM} {SIGSAC} Conference on Computer and
               Communications Security, {CCS} 2022, Los Angeles, CA, USA, November
               7-11, 2022},
  pages     = {2385--2399},
  year      = {2022},
}

@misc{intelproducts,
  title = {Intel® Processors Supporting Intel® SGX},
  author = {Intel(R)},
  url = "https://www.intel.com/content/www/us/en/architecture-and-technology/software-guard-extensions-processors.html",
  year = {2022},
}

@inproceedings{enclavecache,
author = {Chen, Lixia and Li, Jian and Ma, Ruhui and Guan, Haibing and Jacobsen, Hans-Arno},
title = {EnclaveCache: A Secure and Scalable Key-Value Cache in Multi-Tenant Clouds Using Intel SGX},
year = {2019},
isbn = {9781450370097},
publisher = {Association for Computing Machinery},
address = {New York, NY, USA},
url = {https://doi.org/10.1145/3361525.3361533},
doi = {10.1145/3361525.3361533},
pages = {14–27},
numpages = {14},
location = {Davis, CA, USA},
series = {Middleware '19}
}

@inproceedings{aria,
  author    = {Fan Yang and Youmin Chen and Youyou Lu and Qing Wang and Jiwu Shu},
  title     = {Aria: Tolerating Skewed Workloads in Secure In-memory Key-value Stores},
  booktitle = {37th {IEEE} International Conference on Data Engineering ({ICDE})},
  pages     = {1020--1031},
  year      = {2021},
}

@inproceedings{shieldstore,
  author    = {Taehoon Kim and Joongun Park and Jaewook Woo and Seungheun Jeon and Jaehyuk Huh},
  title     = {ShieldStore: Shielded In-memory Key-value Storage with {SGX}},
  booktitle = {Proceedings of the Fourteenth EuroSys Conference},
  year      = {2019},
  pages     = {14:1--14:15},
}

@inproceedings{brandenburger17dsn,
  author    = {Marcus Brandenburger and
               Christian Cachin and
               Matthias Lorenz and
               R{\"{u}}diger Kapitza},
  title     = {Rollback and Forking Detection for Trusted Execution Environments
               Using Lightweight Collective Memory},
  booktitle = {47th Annual {IEEE/IFIP} International Conference on Dependable Systems
               and Networks ({DSN})},
  pages     = {157--168},
  year      = {2017},
}

@inproceedings{strackx16usenix,
  author    = {Raoul Strackx and
               Frank Piessens},
  title     = {Ariadne: {A} Minimal Approach to State Continuity},
  booktitle = {25th {USENIX} Security Symposium},
  pages     = {875--892},
  year      = {2016},
}

@misc{IntelArchitecture,
  title = {Intel(R)64 and IA-32 Architectures Optimization Reference Manual},
  author = {Intel(R)},
  howpublished = {\url{https://software.intel.com/content/www/us/en/develop/download/intel-64-and-ia-32-architectures-optimization-reference-manual.html}},
  year = {2018},
  note = {Accessed: 2020-10-01}
}

@inproceedings{ROTE2017,
author = {Matetic, Sinisa and Ahmed, Mansoor and Kostiainen, Kari and Dhar, Aritra and Sommer, David and Gervais, Arthur and Juels, Ari and Capkun, Srdjan},
title = {ROTE: Rollback Protection for Trusted Execution},
year = {2017},
isbn = {9781931971409},
publisher = {USENIX Association},
address = {USA},
booktitle = {Proceedings of the 26th USENIX Conference on Security Symposium},
pages = {1289–1306},
numpages = {18},
location = {Vancouver, BC, Canada},
series = {SEC'17}
}

@INPROCEEDINGS{EnclaveDB,
  author={C. {Priebe} and K. {Vaswani} and M. {Costa}},
  booktitle={2018 IEEE Symposium on Security and Privacy (SP)},
  title={EnclaveDB: A Secure Database Using SGX},
  year={2018},
  volume={},
  number={},
  pages={264-278},
  doi={10.1109/SP.2018.00025}}

@inproceedings{SGXStep17,
author = {Van Bulck, Jo and Piessens, Frank and Strackx, Raoul},
title = {SGX-Step: A Practical Attack Framework for Precise Enclave Execution Control},
year = {2017},
isbn = {9781450350976},
publisher = {Association for Computing Machinery},
address = {New York, NY, USA},
url = {https://doi.org/10.1145/3152701.3152706},
doi = {10.1145/3152701.3152706},
booktitle = {Proceedings of the 2nd Workshop on System Software for Trusted Execution},
articleno = {4},
numpages = {6},
location = {Shanghai, China},
series = {SysTEX'17}
}

@inproceedings{safekeeper,
  title={Safekeeper: Protecting web passwords using trusted execution environments},
  author={Krawiecka, Klaudia and Kurnikov, Arseny and Paverd, Andrew and Mannan, Mohammad and Asokan, N},
  booktitle={Proceedings of the 2018 World Wide Web Conference},
  pages={349--358},
  year={2018}
}

@article{oblidb,
  title={Oblidb: Oblivious query processing for secure databases},
  author={Eskandarian, Saba and Zaharia, Matei},
  journal={arXiv preprint arXiv:1710.00458},
  year={2017}
}

@article{lind2017teechain,
  title={Teechain: Scalable blockchain payments using trusted execution environments},
  author={Lind, Joshua and Eyal, Ittay and Kelbert, Florian and Naor, Oded and Pietzuch, Peter and Sirer, Emin G{\"u}n},
  journal={arXiv preprint arXiv:1707.05454},
  year={2017}
}

@inproceedings{jangid2021towards,
  title={Towards Formal Verification of State Continuity for Enclave Programs},
  author={Jangid, Mohit Kumar and Chen, Guoxing and Zhang, Yinqian and Lin, Zhiqiang},
  booktitle={30th USENIX Security Symposium (USENIX Security 21)},
  pages={573--590},
  year={2021}
}

@inproceedings {debe,
  author = {Zuoru Yang and Jingwei Li and Patrick P. C. Lee},
  title = {Secure and Lightweight Deduplicated Storage via Shielded {Deduplication-Before-Encryption}},
  booktitle = {2022 USENIX Annual Technical Conference (USENIX ATC 22)},
  year = {2022},
  isbn = {978-1-939133-29-39},
  address = {Carlsbad, CA},
  pages = {37--52},
  url = {https://www.usenix.org/conference/atc22/presentation/yang-zuoru},
  publisher = {USENIX Association},
  month = jul,
}

@misc{awesome_sgx,
  title = {Awesome SGX Open Source Projects},
  year = {2019},
  publisher = {GitHub},
  journal = {GitHub repository},
  howpublished = {\url{https://github.com/Maxul/Awesome-SGX-Open-Source}},
  commit = {955805fc1209b1acd24d227f8bd3729ac0b03b40}
}

@misc{sgx_papers,
  title = {sgx-papers},
  year = {2017},
  publisher = {GitHub},
  journal = {GitHub repository},
  howpublished = {\url{https://github.com/vschiavoni/sgx-papers}},
  commit = {d5382bdb7ae05757916b349595676ee052395195}
}

@inproceedings {twilight_paper,
	author = {Maya Dotan and Saar Tochner and Aviv Zohar and Yossi Gilad},
	title = {Twilight: A Differentially Private Payment Channel Network},
	booktitle = {31st USENIX Security Symposium (USENIX Security 22)},
	year = {2022},
	isbn = {978-1-939133-31-1},
	address = {Boston, MA},
	pages = {555--570},
	url = {https://www.usenix.org/conference/usenixsecurity22/presentation/dotan},
	publisher = {USENIX Association},
	month = aug,
}

@misc{twilight_code,
  title = {Implementation of the paper "Differentially-Private Payment Channels with Twilight"},
  year = {2022},
  publisher = {GitHub},
  journal = {GitHub repository},
  howpublished = {\url{https://github.com/saart/Twilight}},
  commit = {2bc27012f5d0254fc44a73b7fb87cc830b4c5a24}
}

@misc{oasis,
  title = {Oasis Core},
  year = {2018},
  publisher = {GitHub},
  journal = {GitHub repository},
  howpublished = {\url{https://github.com/oasisprotocol/oasis-core}},
  commit = {1cfafd0b20565cbf509dbcd6c3f3bd53c1f2fc82}
}

@misc{mobilecoin_paper,
  author = {},
  title = {Mechanics of MobileCoin: First Edition},
  year = {2021},
  publisher = {MobileCoin Foundation},
  howpublished = {\url{https://mobilecoin.com/learn/read-the-whitepapers/mechanics/}},
  note = {Accessed: 23-02-2023}
}

@misc{mobilecoin_code,
  author = {MobileCoin Foundation},
  title = {MobileCoin},
  year = {2019},
  publisher = {GitHub},
  journal = {GitHub repository},
  howpublished = {\url{https://github.com/mobilecoinfoundation/mobilecoin}},
  commit = {b951090b1efa4a4d9808c8479710d822d6c4ec45}
}

@misc{phala_paper,
  title={Phala network: A confidential smart contract network based on polkadot},
  author={Yin, Hang and Zhou, Shunfan and Jiang, Jun},
  year={2019}
}

@misc{phala_code,
  title = {Phala Blockchain},
  year = {2019},
  publisher = {GitHub},
  journal = {GitHub repository},
  howpublished = {\url{https://github.com/Phala-Network/phala-blockchain}},
  commit = {c5056e01edc0a8747cf6164742bd3636de3cfaef}
}

@inproceedings{teechain_paper,
author = {Lind, Joshua and Naor, Oded and Eyal, Ittay and Kelbert, Florian and Sirer, Emin G\"{u}n and Pietzuch, Peter},
title = {Teechain: A Secure Payment Network with Asynchronous Blockchain Access},
year = {2019},
isbn = {9781450368735},
publisher = {Association for Computing Machinery},
address = {New York, NY, USA},
url = {https://doi.org/10.1145/3341301.3359627},
doi = {10.1145/3341301.3359627},
pages = {63–79},
numpages = {17},
location = {Huntsville, Ontario, Canada},
series = {SOSP '19}
}

@misc{teechain_code,
  author = {Lind, Joshua},
  title = {Teechain: A Secure Payment Network with Asynchronous Blockchain Access},
  year = {2018},
  publisher = {GitHub},
  journal = {GitHub repository},
  howpublished = {\url{https://github.com/lsds/Teechain}},
  commit = {3150fcc91fd908dd34a9305fb154830f567f302c}
}

@INPROCEEDINGS{ekiden_paper,
  author={Cheng, Raymond and Zhang, Fan and Kos, Jernej and He, Warren and Hynes, Nicholas and Johnson, Noah and Juels, Ari and Miller, Andrew and Song, Dawn},
  booktitle={2019 IEEE European Symposium on Security and Privacy (EuroS\&P)},
  title={Ekiden: A Platform for Confidentiality-Preserving, Trustworthy, and Performant Smart Contracts},
  year={2019},
  pages={185-200},
  doi={10.1109/EuroSP.2019.00023}
}

@misc{ekiden_code,
  title = {Ekiden},
  year = {2017},
  publisher = {GitHub},
  journal = {GitHub repository},
  howpublished = {\url{https://github.com/ekiden/ekiden}},
  commit = {2654947f5fcf783c9ca04779af7afdd69622508b}
}

@article{pdo_paper,
  title={Private data objects: an overview},
  author={Bowman, Mic and Miele, Andrea and Steiner, Michael and Vavala, Bruno},
  journal={arXiv preprint arXiv:1807.05686},
  year={2018}
}

@misc{pdo_code,
  author = {Hyperledger Labs},
  title = {Hyperledger Private Data Objects},
  year = {2018},
  publisher = {GitHub},
  journal = {GitHub repository},
  howpublished = {\url{https://github.com/hyperledger-labs/private-data-objects}},
  commit = {cd993a69cd5955ebfe5a9e74b37e26c1b479ddce}
}

@article{fabric_paper,
  title={Blockchain and trusted computing: Problems, pitfalls, and a solution for hyperledger fabric},
  author={Brandenburger, Marcus and Cachin, Christian and Kapitza, R{\"u}diger and Sorniotti, Alessandro},
  journal={arXiv preprint arXiv:1805.08541},
  year={2018}
}

@misc{fabric_code,
  author = {Hyperledger},
  title = {Hyperledger Fabric Private Chaincode},
  year = {2018},
  publisher = {GitHub},
  journal = {GitHub repository},
  howpublished = {\url{https://github.com/hyperledger/fabric-private-chaincode}},
  commit = {a97702902bb6a97475fe4d2170e34569755f03a5}
}

@inproceedings{obscuro_paper,
author = {Tran, Muoi and Luu, Loi and Kang, Min Suk and Bentov, Iddo and Saxena, Prateek},
title = {Obscuro: A Bitcoin Mixer Using Trusted Execution Environments},
year = {2018},
isbn = {9781450365697},
publisher = {Association for Computing Machinery},
address = {New York, NY, USA},
url = {https://doi.org/10.1145/3274694.3274750},
doi = {10.1145/3274694.3274750},
booktitle = {Proceedings of the 34th Annual Computer Security Applications Conference},
pages = {692–701},
numpages = {10},
location = {San Juan, PR, USA},
series = {ACSAC '18}
}

@misc{obscuro_code,
  title = {Obscuro},
  year = {2017},
  publisher = {GitHub},
  journal = {GitHub repository},
  howpublished = {\url{https://github.com/BitObscuro/Obscuro}},
  commit = {101ccf86875cf9b3498894cc6e91a05254c722bb}
}

@inproceedings{pol_paper,
author = {Milutinovic, Mitar and He, Warren and Wu, Howard and Kanwal, Maxinder},
title = {Proof of Luck: An Efficient Blockchain Consensus Protocol},
year = {2016},
isbn = {9781450346702},
publisher = {Association for Computing Machinery},
address = {New York, NY, USA},
url = {https://doi.org/10.1145/3007788.3007790},
doi = {10.1145/3007788.3007790},
booktitle = {Proceedings of the 1st Workshop on System Software for Trusted Execution},
articleno = {2},
numpages = {6},
location = {Trento, Italy},
series = {SysTEX '16}
}

@misc{pol_code,
  author = {},
  title = {Luckychain},
  year = {2016},
  publisher = {GitHub},
  journal = {GitHub repository},
  howpublished = {\url{https://github.com/luckychain/lucky}},
  commit = {d86d7fc4659de13b0c4b224cfb3684177c90eb60}
}

@inproceedings{towncrier_paper,
author = {Zhang, Fan and Cecchetti, Ethan and Croman, Kyle and Juels, Ari and Shi, Elaine},
title = {Town Crier: An Authenticated Data Feed for Smart Contracts},
year = {2016},
isbn = {9781450341394},
publisher = {Association for Computing Machinery},
address = {New York, NY, USA},
url = {https://doi.org/10.1145/2976749.2978326},
doi = {10.1145/2976749.2978326},
booktitle = {Proceedings of the 2016 ACM SIGSAC Conference on Computer and Communications Security},
pages = {270–282},
numpages = {13},
location = {Vienna, Austria},
series = {CCS '16}
}

@misc{towncrier_code,
  author = {},
  title = {Town Crier: An Authenticated Data Feed For Smart Contracts},
  year = {2016},
  publisher = {GitHub},
  journal = {GitHub repository},
  howpublished = {\url{https://github.com/bl4ck5un/Town-Crier}},
  commit = {33471ff56cb75c9672a51c9d9c20352c96cc3444}
}

@INPROCEEDINGS{bloxy,
  author={Rüsch, Signe and Bleeke, Kai and Kapitza, Rüdiger},
  booktitle={2019 38th Symposium on Reliable Distributed Systems (SRDS)},
  title={Bloxy: Providing Transparent and Generic BFT-Based Ordering Services for Blockchains},
  year={2019},
  volume={},
  number={},
  pages={305-30509},
  doi={10.1109/SRDS47363.2019.00043},
  ISSN={2575-8462},
  month={Oct}
}

@inproceedings{troxy,
author = {Behl, Johannes and Distler, Tobias and Kapitza, R\"{u}diger},
title = {Hybrids on Steroids: SGX-Based High Performance BFT},
year = {2017},
isbn = {9781450349383},
publisher = {Association for Computing Machinery},
address = {New York, NY, USA},
url = {https://doi.org/10.1145/3064176.3064213},
doi = {10.1145/3064176.3064213},
booktitle = {Proceedings of the Twelfth European Conference on Computer Systems},
pages = {222–237},
numpages = {16},
location = {Belgrade, Serbia},
series = {EuroSys '17}
}

@inproceedings{bite,
  title={BITE: Bitcoin Lightweight Client Privacy using Trusted Execution.},
  author={Matetic, Sinisa and W{\"u}st, Karl and Schneider, Moritz and Kostiainen, Kari and Karame, Ghassan and Capkun, Srdjan},
  booktitle={USENIX Security Symposium},
  pages={783--800},
  year={2019}
}

@inproceedings {soter_paper,
author = {Tianxiang Shen and Ji Qi and Jianyu Jiang and Xian Wang and Siyuan Wen and Xusheng Chen and Shixiong Zhao and Sen Wang and Li Chen and Xiapu Luo and Fengwei Zhang and Heming Cui},
title = {{SOTER}: Guarding Black-box Inference for General Neural Networks at the Edge},
booktitle = {2022 USENIX Annual Technical Conference (USENIX ATC 22)},
year = {2022},
isbn = {978-1-939133-29-68},
address = {Carlsbad, CA},
pages = {723--738},
url = {https://www.usenix.org/conference/atc22/presentation/shen},
publisher = {USENIX Association},
month = jul,
}

@misc{soter_code,
  author = {},
  title = {Artifact for paper \#1520 SOTER: Guarding Black-box Inference for General Neural Networks at the Edge},
  year = {2022},
  publisher = {GitHub},
  journal = {GitHub repository},
  howpublished = {\url{https://github.com/hku-systems/SOTER}},
  commit = {fb82cdc03f42426c8285d69e9e8a76aea01b85a1}
}

@misc{dp_gbdt,
  author = {Loretan, Rudolf},
  title = {Enclave hardening for private ML},
  year = {2021},
  publisher = {GitHub},
  journal = {GitHub repository},
  howpublished = {\url{https://github.com/loretanr/dp-gbdt}},
  commit = {1685641b604ac0f19e4d160e15975326d48b36f8}
}

@inproceedings{xgboost_paper,
author = {Law, Andrew and Leung, Chester and Poddar, Rishabh and Popa, Raluca Ada and Shi, Chenyu and Sima, Octavian and Yu, Chaofan and Zhang, Xingmeng and Zheng, Wenting},
title = {Secure Collaborative Training and Inference for XGBoost},
year = {2020},
isbn = {9781450380881},
publisher = {Association for Computing Machinery},
address = {New York, NY, USA},
url = {https://doi.org/10.1145/3411501.3419420},
doi = {10.1145/3411501.3419420},
booktitle = {Proceedings of the 2020 Workshop on Privacy-Preserving Machine Learning in Practice},
pages = {21–26},
numpages = {6},
location = {Virtual Event, USA},
series = {PPMLP'20}
}

@misc{xgboost_code,
  author = {},
  title = {Secure XGBoost},
  year = {2020},
  publisher = {GitHub},
  journal = {GitHub repository},
  howpublished = {\url{https://github.com/mc2-project/secure-xgboost}},
  commit = {03148e31b52641d519dd97a823c9cc3a84a35e24}
}

@misc{confidential_ml,
  author = {Jose, Prasad Koshy},
  title = {Confidential Computing of Machine Learning using Intel SGX},
  year = {2020},
  publisher = {GitHub},
  journal = {GitHub repository},
  howpublished = {\url{https://github.com/prasadkjose/confidential-ml-sgx}},
  commit = {6de9b124cd812c609822b063dfab42a708cd3b28}
}

@article{slalom_paper,
  title={Slalom: Fast, verifiable and private execution of neural networks in trusted hardware},
  author={Tramer, Florian and Boneh, Dan},
  journal={arXiv preprint arXiv:1806.03287},
  year={2018}
}

@misc{slalom_code,
  author = {Tramèr, Florian and Boneh, Dan},
  title = {SLALOM},
  year = {2018},
  publisher = {GitHub},
  journal = {GitHub repository},
  howpublished = {\url{https://github.com/ftramer/slalom}},
  commit = {96ff15977b7058b96d2a00a51c6aabbe729cc6d5}
}

@INPROCEEDINGS{plinius_paper,
  author={Yuhala, Peterson and Felber, Pascal and Schiavoni, Valerio and Tchana, Alain},
  booktitle={2021 51st Annual IEEE/IFIP International Conference on Dependable Systems and Networks (DSN)},
  title={Plinius: Secure and Persistent Machine Learning Model Training},
  year={2021},
  volume={},
  number={},
  pages={52-62},
  doi={10.1109/DSN48987.2021.00022},
  ISSN={2158-3927},
  month={June},
}

@misc{plinius_code,
  author = {},
  title = {Plinius},
  year = {2020},
  publisher = {GitHub},
  journal = {GitHub repository},
  howpublished = {\url{https://github.com/anonymous-xh/plinius}},
  commit = {a251629a05a15c4db545ab8e7a19f790c07a5ca3}
}

@inproceedings{securetf,
author = {Quoc, Do Le and Gregor, Franz and Arnautov, Sergei and Kunkel, Roland and Bhatotia, Pramod and Fetzer, Christof},
title = {SecureTF: A Secure TensorFlow Framework},
year = {2020},
isbn = {9781450381536},
publisher = {Association for Computing Machinery},
address = {New York, NY, USA},
url = {https://doi.org/10.1145/3423211.3425687},
doi = {10.1145/3423211.3425687},
booktitle = {Proceedings of the 21st International Middleware Conference},
pages = {44–59},
numpages = {16},
location = {Delft, Netherlands},
series = {Middleware '20}
}

@article{cacic_paper,
author = {Thomaz, Guilherme A. and Guerra, Matheus B. and Sammarco, Matteo and Detyniecki, Marcin and Campista, Miguel Elias M.},
title = {Tamper-proof Access Control for IoT Clouds Using Enclaves},
year = {2022},
isbn = {},
publisher = {Association for Computing Machinery},
url = {https://www.gta.ufrj.br/ftp/gta/TechReports/TGS23.pdf},
}

@misc{cacic_code,
  author = {},
  title = {CACIC Use Case},
  year = {2023},
  publisher = {GitHub},
  journal = {GitHub repository},
  howpublished = {\url{https://github.com/GTA-UFRJ/CACIC-Use-Case}},
  commit = {98967939958466d819ef44997dbb3b5bf59b71f5}
}

@inproceedings {debe_paper,
author = {Zuoru Yang and Jingwei Li and Patrick P. C. Lee},
title = {Secure and Lightweight Deduplicated Storage via Shielded {Deduplication-Before-Encryption}},
booktitle = {2022 USENIX Annual Technical Conference (USENIX ATC 22)},
year = {2022},
isbn = {978-1-939133-29-39},
address = {Carlsbad, CA},
pages = {37--52},
url = {https://www.usenix.org/conference/atc22/presentation/yang-zuoru},
publisher = {USENIX Association},
month = jul,
}

@misc{debe_code,
  author = {},
  title = {DEBE},
  year = {2022},
  publisher = {GitHub},
  journal = {GitHub repository},
  howpublished = {\url{https://github.com/yzr95924/DEBE}},
  commit = {2e62036563d9c27d8d65fab22d89d265a7e79f7f}
}

@INPROCEEDINGS{rex_paper,
  author={Dhasade, Akash and Dresevic, Nevena and Kermarrec, Anne-Marie and Pires, Rafael},
  booktitle={2022 IEEE International Parallel and Distributed Processing Symposium (IPDPS)},
  title={{TEE}-based decentralized recommender systems: The raw data sharing redemption},
  year={2022},
  volume={},
  number={},
  pages={447-458},
  doi={10.1109/IPDPS53621.2022.00050}
}

@misc{rex_code,
  author = {},
  title = {{REX}: SGX decentralized recommender},
  year = {2022},
  publisher = {GitHub},
  journal = {GitHub repository},
  howpublished = {\url{https://github.com/rafaelppires/rex}},
  commit = {e2003d0fdc24329d89947ba23db0830ab18940da}
}

@inproceedings{dedup_paper,
  title={SGXDedup},
  author={Ren, Yanjing and Li, Jingwei and Yang, Zuoru and Lee, Patrick PC and Zhang, Xiaosong},
  booktitle={USENIX Annual Technical Conference},
  pages={957--971},
  year={2021}
}

@misc{dedup_code,
  author = {},
  title = {Accelerating Encrypted Deduplication via SGX},
  year = {2021},
  publisher = {GitHub},
  journal = {GitHub repository},
  howpublished = {\url{https://github.com/jingwei87/sgxdedup}},
  commit = {e1e9bc2a3755fa79b6477d9a4c23928dc1d80af4}
}

@article{skses_paper,
  title={Sketching algorithms for genomic data analysis and querying in a secure enclave},
  author={Kockan, Can and Zhu, Kaiyuan and Dokmai, Natnatee and Karpov, Nikolai and Kulekci, M Oguzhan and Woodruff, David P and Sahinalp, S Cenk},
  journal={Nature methods},
  volume={17},
  number={3},
  pages={295--301},
  year={2020},
  publisher={Nature Publishing Group US New York}
}

@misc{skses_code,
  author = {},
  title = {SkSES},
  year = {2018},
  publisher = {GitHub},
  journal = {GitHub repository},
  howpublished = {\url{https://github.com/ndokmai/sgx-genome-variants-search}},
  commit = {dd83fb53d0a82594b9ab2c253a246a80095ca12b}
}

@article{smac_gen_paper,
  title={Privacy-preserving genotype imputation in a trusted execution environment},
  author={Dokmai, Natnatee and Kockan, Can and Zhu, Kaiyuan and Wang, XiaoFeng and Sahinalp, S Cenk and Cho, Hyunghoon},
  journal={Cell systems},
  volume={12},
  number={10},
  pages={983--993},
  year={2021},
  publisher={Elsevier}
}

@misc{smac_gen_code,
  author = {},
  title = {SMac: Secure Genotype Imputation in Intel SGX},
  year = {2020},
  publisher = {GitHub},
  journal = {GitHub repository},
  howpublished = {\url{https://github.com/ndokmai/sgx-genotype-imputation}},
  commit = {c2520881a74d94a6cd9f8ee7278d87a112229912}
}

@INPROCEEDINGS{hysec_paper,
  author={Widanage, Chathura and Liu, Weijie and Li, Jiayu and Chen, Hongbo and Wang, XiaoFeng and Tang, Haixu and Fox, Judy},
  booktitle={2021 IEEE 14th International Conference on Cloud Computing (CLOUD)},
  title={HySec-Flow: Privacy-Preserving Genomic Computing with SGX-based Big-Data Analytics Framework},
  year={2021},
  volume={},
  number={},
  pages={733-743},
  doi={10.1109/CLOUD53861.2021.00098}}

@misc{hysec_code,
  author = {},
  title = {bwa-sgx-scone},
  year = {2021},
  publisher = {GitHub},
  journal = {GitHub repository},
  howpublished = {\url{https://github.com/dsc-sgx/bwa-sgx-scone}},
  commit = {a90b307a5051198d3c6181d77aecd112f9d2ba3e}
}

@misc{bisgx_code,
  author = {},
  title = {BI-SGX : Bioinformatic Interpreter on SGX-based Secure Computing Cloud},
  year = {2018},
  publisher = {GitHub},
  journal = {GitHub repository},
  howpublished = {\url{https://github.com/hello31337/BI-SGX}},
  commit = {bce261dc243355e855da5fcac857a1ff43627c60}
}

@article{signal_blog,
  title={Technology preview: Private contact discovery for Signal},
  author={Marlinspike, Moxie},
  year={2017},
  publisher={Signal Blog},
  howpublished = {\url{https://signal.org/blog/private-contact-discovery/}},
  note = {Accessed: 09-03-2023}
}

@misc{signal_code,
  author = {},
  title = {Private Contact Discovery Service (Beta)},
  year = {2017},
  publisher = {GitHub},
  journal = {GitHub repository},
  howpublished = {\url{https://github.com/signalapp/ContactDiscoveryService}},
  commit = {3ee0ca31a2e2fd2568b6ed9297093245cdd4deda}
}

@inproceedings{tresorsgx_paper,
author = {Richter, Lars and G\"{o}tzfried, Johannes and M\"{u}ller, Tilo},
title = {Isolating Operating System Components with Intel SGX},
year = {2016},
isbn = {9781450346702},
publisher = {Association for Computing Machinery},
address = {New York, NY, USA},
url = {https://doi.org/10.1145/3007788.3007796},
doi = {10.1145/3007788.3007796},
booktitle = {Proceedings of the 1st Workshop on System Software for Trusted Execution},
articleno = {8},
numpages = {6},
location = {Trento, Italy},
series = {SysTEX '16}
}

@misc{tresorsgx_code,
  author = {},
  title = {TresorSGX},
  year = {2016},
  publisher = {GitHub},
  journal = {GitHub repository},
  howpublished = {\url{https://github.com/ayeks/TresorSGX}},
  commit = {d2e529ea977fef4042f591be5bb5fdfccbee1df4}
}

@inproceedings{privatube,
	author = {Da Silva, Simon and Ben Mokhtar, Sonia and Contiu, Stefan and N\'{e}gru, Daniel and R\'{e}veill\`{e}re, Laurent and Rivi\`{e}re, Etienne},
	title = {PrivaTube: Privacy-Preserving Edge-Assisted Video Streaming},
	year = {2019},
	isbn = {9781450370097},
	publisher = {Association for Computing Machinery},
	address = {New York, NY, USA},
	url = {https://doi.org/10.1145/3361525.3361546},
	doi = {10.1145/3361525.3361546},
	booktitle = {Proceedings of the 20th International Middleware Conference},
	pages = {189–201},
	numpages = {13},
	location = {Davis, CA, USA},
	series = {Middleware '19}
}

@inproceedings{v2v_paper,
  title={Towards a TEE-based V2V Protocol for Connected and Autonomous Vehicles},
  author={Jangid, M and Lin, Zhiqiang},
  booktitle={Workshop on Automotive and Autonomous Vehicle Security (AutoSec)},
  year={2022}
}

@misc{v2v_code,
  author = {},
  title = {V2V SGX},
  year = {2022},
  publisher = {GitHub},
  journal = {GitHub repository},
  howpublished = {\url{https://github.com/OSUSecLab/v2v-sgx-prelim}},
  commit = {575109dd587abe80b3069c1f1d6596ab35feea4a}
}

@article{macsec,
  title={Secure Network Interface with SGX},
  author={Kirchengast, Felix},
  year={2019},
  publisher={GitHub},
  journal = {GitHub repository},
  howpublished = {\url{https://github.com/fkirc/secure-network-interface-with-sgx}},
  commit = {396175603b473fb619c39d20753df3fccbe83d61}
}

@inproceedings{seng_paper,
  title={SENG, the sgx-enforcing network gateway: Authorizing communication from shielded clients},
  author={Schwarz, Fabian and Rossow, Christian},
  booktitle={Proceedings of the 29th USENIX Conference on Security Symposium},
  pages={753--770},
  year={2020}
}

@misc{seng_code,
  author = {},
  title = {SENG, the SGX-Enforcing Network Gateway},
  year = {2020},
  publisher = {GitHub},
  journal = {GitHub repository},
  howpublished = {\url{https://github.com/sengsgx/sengsgx}},
  commit = {1f0c0262bea3d61a3320ba2beb3288a85c701cc4}
}

@article{consensgx_paper,
  title={ConsenSGX: Scaling Anonymous Communications Networks with Trusted Execution Environments.},
  author={Sasy, Sajin and Goldberg, Ian},
  journal={Proc. Priv. Enhancing Technol.},
  volume={2019},
  number={3},
  pages={331--349},
  year={2019}
}

@misc{consensgx_code,
  author = {},
  title = {ConsenSGX},
  year = {2019},
  publisher = {GitHub},
  journal = {GitHub repository},
  howpublished = {\url{https://github.com/sshsshy/ConsenSGX}},
  commit = {b5267c5c194c4d53c4d0c3fec35bda93500e725f}
}

@inproceedings{safebricks_paper,
  title={Safebricks: Shielding network functions in the cloud},
  author={Poddar, Rishabh and Lan, Chang and Popa, Raluca Ada and Ratnasamy, Sylvia},
  booktitle={15th $\{$USENIX$\}$ Symposium on Networked Systems Design and Implementation ($\{$NSDI$\}$ 18)},
  pages={201--216},
  year={2018}
}

@misc{safebricks_code,
  author = {},
  title = {SafeBricks},
  year = {2016},
  publisher = {GitHub},
  journal = {GitHub repository},
  howpublished = {\url{https://github.com/YangZhou1997/SafeBricks}},
  commit = {cd81bbe60264a983b37f1aa6fde91fae89d7e4ad}
}

@ARTICLE{sgxtor_paper,
  author={Kim, Seongmin and Han, Juhyeng and Ha, Jaehyeong and Kim, Taesoo and Han, Dongsu},
  journal={IEEE/ACM Transactions on Networking},
  title={SGX-Tor: A Secure and Practical Tor Anonymity Network With SGX Enclaves},
  year={2018},
  volume={26},
  number={5},
  pages={2174-2187},
  doi={10.1109/TNET.2018.2868054}}

@misc{sgxtor_code,
  author = {},
  title = {SGX-Tor},
  year = {2017},
  publisher = {GitHub},
  journal = {GitHub repository},
  howpublished = {\url{https://github.com/kaist-ina/SGX-Tor}},
  commit = {193d4f072d49799a25830c75ef7b29f0f960e66d}
}

@inproceedings{sgxcbr,
author = {Pires, Rafael and Pasin, Marcelo and Felber, Pascal and Fetzer, Christof},
title = {Secure Content-Based Routing Using Intel Software Guard Extensions},
year = {2016},
isbn = {9781450343008},
publisher = {Association for Computing Machinery},
address = {New York, NY, USA},
url = {https://doi.org/10.1145/2988336.2988346},
doi = {10.1145/2988336.2988346},
booktitle = {Proceedings of the 17th International Middleware Conference},
articleno = {10},
numpages = {10},
location = {Trento, Italy},
series = {Middleware '16}
}

@INPROCEEDINGS{cyclosa,
  author={Pires, Rafael and Goltzsche, David and Ben Mokhtar, Sonia and Bouchenak, Sara and Boutet, Antoine and Felber, Pascal and Kapitza, Rüdiger and Pasin, Marcelo and Schiavoni, Valerio},
  booktitle={2018 IEEE 38th International Conference on Distributed Computing Systems (ICDCS)},
  title={CYCLOSA: Decentralizing Private Web Search through SGX-Based Browser Extensions},
  year={2018},
  volume={},
  number={},
  pages={467-477},
  doi={10.1109/ICDCS.2018.00053}
}

@inproceedings{snfv,
author = {Shih, Ming-Wei and Kumar, Mohan and Kim, Taesoo and Gavrilovska, Ada},
title = {S-NFV: Securing NFV States by Using SGX},
year = {2016},
isbn = {9781450340786},
publisher = {Association for Computing Machinery},
address = {New York, NY, USA},
url = {https://doi.org/10.1145/2876019.2876032},
doi = {10.1145/2876019.2876032},
booktitle = {Proceedings of the 2016 ACM International Workshop on Security in Software Defined Networks \& Network Function Virtualization},
pages = {45–48},
numpages = {4},
location = {New Orleans, Louisiana, USA},
series = {SDN-NFV Security '16}
}

@INPROCEEDINGS{pubsub_paper,
  author={Arnautov, Sergei and Brito, Andrey and Felber, Pascal and Fetzer, Christof and Gregor, Franz and Krahn, Robert and Ozga, Wojciech and Martin, André and Schiavoni, Valerio and Silva, Fábio and Tenorio, Marcus and Thümmel, Nikolaus},
  booktitle={2018 IEEE 37th Symposium on Reliable Distributed Systems (SRDS)},
  title={PubSub-SGX: Exploiting Trusted Execution Environments for Privacy-Preserving Publish/Subscribe Systems},
  year={2018},
  volume={},
  number={},
  pages={123-132},
  doi={10.1109/SRDS.2018.00023}
}

@misc{pubsub_code,
  author = {},
  title = {The SELIS Publish/Subscribe system},
  year = {2019},
  publisher = {GitHub},
  journal = {GitHub repository},
  howpublished = {\url{https://github.com/selisproject/pubsub}},
  commit = {3650f52887055d17816dbe0939e96142cb9c8f85}
}

@INPROCEEDINGS{endbox,
  author={Goltzsche, David and Rüsch, Signe and Nieke, Manuel and Vaucher, Sébastien and Weichbrodt, Nico and Schiavoni, Valerio and Aublin, Pierre-Louis and Cosa, Paolo and Fetzer, Christof and Felber, Pascal and Pietzuch, Peter and Kapitza, Rüdiger},
  booktitle={2018 48th Annual IEEE/IFIP International Conference on Dependable Systems and Networks (DSN)},
  title={EndBox: Scalable Middlebox Functions Using Client-Side Trusted Execution},
  year={2018},
  volume={},
  number={},
  pages={386-397},
  doi={10.1109/DSN.2018.00048}
}

@inproceedings{lightbox_paper,
author = {Duan, Huayi and Wang, Cong and Yuan, Xingliang and Zhou, Yajin and Wang, Qian and Ren, Kui},
title = {LightBox: Full-Stack Protected Stateful Middlebox at Lightning Speed},
year = {2019},
isbn = {9781450367479},
publisher = {Association for Computing Machinery},
address = {New York, NY, USA},
url = {https://doi.org/10.1145/3319535.3339814},
doi = {10.1145/3319535.3339814},
booktitle = {Proceedings of the 2019 ACM SIGSAC Conference on Computer and Communications Security},
pages = {2351–2367},
numpages = {17},
location = {London, United Kingdom},
series = {CCS '19}
}

@misc{lightbox_code,
  author = {},
  title = {LightBox},
  year = {2018},
  publisher = {GitHub},
  journal = {GitHub repository},
  howpublished = {\url{https://github.com/lightbox-impl/LightBox}},
  commit = {eb7c1f1139917731623aa17af643fae2a6beccf4}
}

@inproceedings{opaque_paper,
  title={Opaque: An Oblivious and Encrypted Distributed Analytics Platform.},
  author={Zheng, Wenting and Dave, Ankur and Beekman, Jethro G and Popa, Raluca Ada and Gonzalez, Joseph E and Stoica, Ion},
  booktitle={NSDI},
  volume={17},
  pages={283--298},
  year={2017}
}

@misc{opaque_code,
  author = {},
  title = {Opaque},
  year = {2017},
  publisher = {GitHub},
  journal = {GitHub repository},
  howpublished = {\url{https://github.com/mc2-project/opaque-sql}},
  commit = {d0bb5a9c1925945c0562a21051f8c079aad418ad}
}

@inproceedings{snoopy_paper,
author = {Dauterman, Emma and Fang, Vivian and Demertzis, Ioannis and Crooks, Natacha and Popa, Raluca Ada},
title = {Snoopy: Surpassing the Scalability Bottleneck of Oblivious Storage},
year = {2021},
isbn = {9781450387095},
publisher = {Association for Computing Machinery},
address = {New York, NY, USA},
url = {https://doi.org/10.1145/3477132.3483562},
doi = {10.1145/3477132.3483562},
booktitle = {Proceedings of the ACM SIGOPS 28th Symposium on Operating Systems Principles},
pages = {655–671},
numpages = {17},
location = {Virtual Event, Germany},
series = {SOSP '21}
}

@misc{snoopy_code,
  author = {},
  title = {Snoopy: A Scalable Oblivious Storage System},
  year = {2021},
  publisher = {GitHub},
  journal = {GitHub repository},
  howpublished = {\url{https://github.com/ucbrise/snoopy}},
  commit = {da4c98e3876c10cf52aa51ece3b62c5e8b8e335a}
}

@inproceedings{desearch_paper,
  title={Bringing Decentralized Search to Decentralized Services.},
  author={Li, Mingyu and Zhu, Jinhao and Zhang, Tianxu and Tan, Cheng and Xia, Yubin and Angel, Sebastian and Chen, Haibo},
  booktitle={OSDI},
  pages={331--347},
  year={2021}
}

@misc{desearch_code,
  author = {},
  title = {Desearch},
  year = {2021},
  publisher = {GitHub},
  journal = {GitHub repository},
  howpublished = {\url{https://github.com/SJTU-IPADS/DeSearch}},
  commit = {92ebd3cb7ea832243112968add39283d0e844f9d}
}

@inproceedings{xsearch_paper,
author = {Mokhtar, Sonia Ben and Boutet, Antoine and Felber, Pascal and Pasin, Marcelo and Pires, Rafael and Schiavoni, Valerio},
title = {X-Search: Revisiting Private Web Search Using Intel SGX},
year = {2017},
isbn = {9781450347204},
publisher = {Association for Computing Machinery},
address = {New York, NY, USA},
url = {https://doi.org/10.1145/3135974.3135987},
doi = {10.1145/3135974.3135987},
booktitle = {Proceedings of the 18th ACM/IFIP/USENIX Middleware Conference},
pages = {198–208},
numpages = {11},
location = {Las Vegas, Nevada},
series = {Middleware '17}
}

@misc{xsearch_code,
  author = {},
  title = {X-Search},
  year = {2020},
  publisher = {GitHub},
  journal = {GitHub repository},
  howpublished = {\url{https://github.com/Sand-jrd/SGX-Search}},
  commit = {4d065ad2f8182d57e14a4ba90de3eb2c18f4e420}
}

@InProceedings{maiden_paper,
author="Vo, Viet
and Lai, Shangqi
and Yuan, Xingliang
and Nepal, Surya
and Liu, Joseph K.",
title="Towards Efficient and Strong Backward Private Searchable Encryption with Secure Enclaves",
booktitle="Applied Cryptography and Network Security",
year="2021",
publisher="Springer International Publishing",
address="Cham",
pages="50--75",
isbn="978-3-030-78372-3"
}

@misc{maiden_code,
  author = {},
  title = {SGXSSE Maiden},
  year = {2020},
  publisher = {GitHub},
  journal = {GitHub repository},
  howpublished = {\url{https://github.com/MonashCybersecurityLab/SGXSSE}},
  commit = {052069563303c6b2fa8dc3069cef450b678e2ea3}
}

@article{posup_paper,
  title={Hardware-supported ORAM in effect: Practical oblivious search and update on very large dataset},
  author={Hoang, Thang and Ozmen, Muslum Ozgur and Jang, Yeongjin and Yavuz, Attila A},
  journal={Proceedings on Privacy Enhancing Technologies},
  volume={2019},
  number={1},
  pages={172--191},
  year={2019}
}

@misc{posup_code,
  author = {},
  title = {POSUP: Oblivious Search and Update Platform with SGX},
  year = {2018},
  publisher = {GitHub},
  journal = {GitHub repository},
  howpublished = {\url{https://github.com/thanghoang/POSUP}},
  commit = {9a6e181763535fd927bca1749cd804e5d77f2f73}
}

@ARTICLE{qshield_paper,
  author={Chen, Yaxing and Zheng, Qinghua and Yan, Zheng and Liu, Dan},
  journal={IEEE Transactions on Parallel and Distributed Systems},
  title={QShield: Protecting Outsourced Cloud Data Queries With Multi-User Access Control Based on SGX},
  year={2021},
  volume={32},
  number={2},
  pages={485-499},
  doi={10.1109/TPDS.2020.3024880}
}

@misc{qshield_code,
  author = {},
  title = {QShield},
  year = {2020},
  publisher = {GitHub},
  journal = {GitHub repository},
  howpublished = {\url{https://github.com/fishermano/QShield}},
  commit = {0eb4359f826a891ee77a3e3421c1e290a25fee53}
}

@ARTICLE{bisen_paper,
  author={Ferreira, Bernardo and Portela, Bernardo and Oliveira, Tiago and Borges, Guilherme and Domingos, Henrique and Leitão, João},
  journal={IEEE Transactions on Dependable and Secure Computing},
  title={Boolean Searchable Symmetric Encryption With Filters on Trusted Hardware},
  year={2022},
  volume={19},
  number={2},
  pages={1307-1319},
  doi={10.1109/TDSC.2020.3012100}
}

@misc{bisen_code,
  author = {},
  title = {Boolean Isolated Searchable Encryption (BISEN)},
  year = {2019},
  publisher = {GitHub},
  journal = {GitHub repository},
  howpublished = {\url{https://github.com/bernymac/BISEN}},
  commit = {b5c922dfdd1c842f26ad6e249f06899f807fba4b}
}

@inproceedings{feido_paper,
author = {Schwarz, Fabian and Do, Khue and Heide, Gunnar and Hanzlik, Lucjan and Rossow, Christian},
title = {FeIDo: Recoverable FIDO2 Tokens Using Electronic IDs},
year = {2022},
isbn = {9781450394505},
publisher = {Association for Computing Machinery},
address = {New York, NY, USA},
url = {https://doi.org/10.1145/3548606.3560584},
doi = {10.1145/3548606.3560584},
booktitle = {Proceedings of the 2022 ACM SIGSAC Conference on Computer and Communications Security},
pages = {2581–2594},
numpages = {14},
location = {Los Angeles, CA, USA},
series = {CCS '22}
}

@misc{feido_code,
  author = {},
  title = {FeIDo Credential Service, Intel SGX version},
  year = {2022},
  publisher = {GitHub},
  journal = {GitHub repository},
  howpublished = {\url{https://github.com/feido-token}},
  commit = {4a2710a4591d4170d042defd1784f0196727820e}
}

@article{sgxkms_paper,
  title={Intel SGX enabled key manager service with openstack barbican},
  author={Chakrabarti, Somnath and Baker, Brandon and Vij, Mona},
  journal={arXiv preprint arXiv:1712.07694},
  year={2017}
}

@misc{sgxkms_code,
  author = {},
  title = {SGX Enabled OpenStack Barbican Key Management System},
  year = {2017},
  publisher = {GitHub},
  journal = {GitHub repository},
  howpublished = {\url{https://github.com/cloud-security-research/sgx-kms}},
  commit = {1c9d2d2df117845166196a1c3f6f2fb1b998e2c9}
}

@inproceedings{keyscloud_paper,
author = {Kurnikov, Arseny and Paverd, Andrew and Mannan, Mohammad and Asokan, N.},
title = {Keys in the Clouds: Auditable Multi-Device Access to Cryptographic Credentials},
year = {2018},
isbn = {9781450364485},
publisher = {Association for Computing Machinery},
address = {New York, NY, USA},
url = {https://doi.org/10.1145/3230833.3234518},
doi = {10.1145/3230833.3234518},
booktitle = {Proceedings of the 13th International Conference on Availability, Reliability and Security},
articleno = {40},
numpages = {10},
location = {Hamburg, Germany},
series = {ARES 2018}
}

@misc{keyscloud_code,
  author = {},
  title = {Cloud Key Store - secure storage for private credentials},
  year = {2018},
  publisher = {GitHub},
  journal = {GitHub repository},
  howpublished = {\url{https://github.com/cloud-key-store/keystore}},
  commit = {0832cdadd29a2438414fbdbbd361d32545ccd02f}
}

@inproceedings{safekeeper_paper,
author = {Krawiecka, Klaudia and Kurnikov, Arseny and Paverd, Andrew and Mannan, Mohammad and Asokan, N.},
title = {SafeKeeper: Protecting Web Passwords Using Trusted Execution Environments},
year = {2018},
isbn = {9781450356398},
publisher = {International World Wide Web Conferences Steering Committee},
address = {Republic and Canton of Geneva, CHE},
url = {https://doi.org/10.1145/3178876.3186101},
doi = {10.1145/3178876.3186101},
booktitle = {Proceedings of the 2018 World Wide Web Conference},
pages = {349–358},
numpages = {10},
location = {Lyon, France},
series = {WWW '18}
}

@misc{safekeeper_code,
  author = {},
  title = {SafeKeeper - Protecting Web passwords using Trusted Execution Environments},
  year = {2018},
  publisher = {GitHub},
  journal = {GitHub repository},
  howpublished = {\url{https://github.com/SafeKeeper/safekeeper-server}},
  commit = {9c7a479208f4e27cb6f60bbd8b00a03b9544222d}
}

@inproceedings{delegatee,
  title={DelegaTEE: Brokered Delegation Using Trusted Execution Environments.},
  author={Matetic, Sinisa and Schneider, Moritz and Miller, Andrew and Juels, Ari and Capkun, Srdjan},
  booktitle={USENIX Security Symposium},
  pages={1387--1403},
  year={2018}
}

@inproceedings{avocado_paper,
  title={Avocado: A Secure In-Memory Distributed Storage System.},
  author={Bailleu, Maurice and Giantsidi, Dimitra and Gavrielatos, Vasilis and Do Le Quoc and Nagarajan, Vijay and Bhatotia, Pramod},
  booktitle={USENIX Annual Technical Conference},
  pages={65--79},
  year={2021}
}

@misc{avocado_code,
  author = {},
  title = {Avocado},
  year = {2021},
  publisher = {GitHub},
  journal = {GitHub repository},
  howpublished = {\url{https://github.com/mbailleu/avocado}},
  commit = {6b835f580671412be975fea70b8d1890cdd8e399}
}

@inproceedings{speicher_paper,
  title={SPEICHER: Securing LSM-based Key-Value Stores using Shielded Execution.},
  author={Bailleu, Maurice and Thalheim, J{\"o}rg and Bhatotia, Pramod and Fetzer, Christof and Honda, Michio and Vaswani, Kapil},
  booktitle={FAST},
  pages={173--190},
  year={2019}
}

@misc{speicher_code,
  author = {},
  title = {SpeicherDPDK},
  year = {2015},
  publisher = {GitHub},
  journal = {GitHub repository},
  howpublished = {\url{https://github.com/mbailleu/SpeicherDPDK}},
  commit = {e1b55cb818566857d704aada220b956098e99e7d}
}

@article{stealthdb_paper,
  title={StealthDB: a Scalable Encrypted Database with Full SQL Query Support.},
  author={Vinayagamurthy, Dhinakaran and Gribov, Alexey and Gorbunov, Sergey},
  journal={Proc. Priv. Enhancing Technol.},
  volume={2019},
  number={3},
  pages={370--388},
  year={2019}
}

@misc{stealthdb_code,
  author = {},
  title = {StealthDB},
  year = {2017},
  publisher = {GitHub},
  journal = {GitHub repository},
  howpublished = {\url{https://github.com/cryptograph/stealthdb}},
  commit = {1ca645ae1613c146d59900ce50abc873dc8a6d01}
}

@INPROCEEDINGS{stanlite_paper,
  author={Sartakov, Vasily and Weichbrodt, Nico and Krieter, Sebastian and Leich, Thomas and Kapitza, Rudiger},
  booktitle={2018 IEEE International Conference on Cloud Engineering (IC2E)},
  title={STANlite – A Database Engine for Secure Data Processing at Rack-Scale Level},
  year={2018},
  volume={},
  number={},
  pages={23-33},
  doi={10.1109/IC2E.2018.00024}
}

@misc{stanlite_code,
  author = {},
  title = {STANlite},
  year = {2020},
  publisher = {GitHub},
  journal = {GitHub repository},
  howpublished = {\url{https://github.com/ibr-ds/STANlite}},
  commit = {16467c8034e84bd9a1cea611e126187adedcad74}
}

@inproceedings{shieldstore_paper,
author = {Kim, Taehoon and Park, Joongun and Woo, Jaewook and Jeon, Seungheun and Huh, Jaehyuk},
title = {ShieldStore: Shielded In-Memory Key-Value Storage with SGX},
year = {2019},
isbn = {9781450362818},
publisher = {Association for Computing Machinery},
address = {New York, NY, USA},
url = {https://doi.org/10.1145/3302424.3303951},
doi = {10.1145/3302424.3303951},
booktitle = {Proceedings of the Fourteenth EuroSys Conference 2019},
articleno = {14},
numpages = {15},
location = {Dresden, Germany},
series = {EuroSys '19}
}

@misc{shieldstore_code,
  author = {},
  title = {ShieldStore},
  year = {2018},
  publisher = {GitHub},
  journal = {GitHub repository},
  howpublished = {\url{https://github.com/cocoppang/ShieldStore}},
  commit = {5de0122c1281730271649c7bd0980d496903698b}
}

@article{oblidb_paper,
  title={Oblidb: Oblivious query processing for secure databases},
  author={Eskandarian, Saba and Zaharia, Matei},
  journal={arXiv preprint arXiv:1710.00458},
  year={2017}
}

@misc{oblidb_code,
  author = {},
  title = {ObliDB},
  year = {2017},
  publisher = {GitHub},
  journal = {GitHub repository},
  howpublished = {\url{https://github.com/SabaEskandarian/ObliDB}},
  commit = {6de5ba15eb27b4ddc86481e2eb8db549159344ea}
}

@InProceedings{hardidx,
author="Fuhry, Benny
and Bahmani, Raad
and Brasser, Ferdinand
and Hahn, Florian
and Kerschbaum, Florian
and Sadeghi, Ahmad-Reza",
title="HardIDX: Practical and Secure Index with SGX",
booktitle="Data and Applications Security and Privacy XXXI",
year="2017",
publisher="Springer International Publishing",
address="Cham",
pages="386--408",
isbn="978-3-319-61176-1"
}

@inproceedings{pesos,
author = {Krahn, Robert and Trach, Bohdan and Vahldiek-Oberwagner, Anjo and Knauth, Thomas and Bhatotia, Pramod and Fetzer, Christof},
title = {Pesos: Policy Enhanced Secure Object Store},
year = {2018},
isbn = {9781450355841},
publisher = {Association for Computing Machinery},
address = {New York, NY, USA},
url = {https://doi.org/10.1145/3190508.3190518},
doi = {10.1145/3190508.3190518},
booktitle = {Proceedings of the Thirteenth EuroSys Conference},
articleno = {25},
numpages = {17},
location = {Porto, Portugal},
series = {EuroSys '18}
}

@INPROCEEDINGS{nexus_paper,
  author={Djoko, Judicael B. and Lange, Jack and Lee, Adam J.},
  booktitle={2019 49th Annual IEEE/IFIP International Conference on Dependable Systems and Networks (DSN)},
  title={NeXUS: Practical and Secure Access Control on Untrusted Storage Platforms using Client-Side SGX},
  year={2019},
  volume={},
  number={},
  pages={401-413},
  doi={10.1109/DSN.2019.00049}
}

@misc{nexus_code,
  author = {},
  title = {NeXUS},
  year = {2017},
  publisher = {GitHub},
  journal = {GitHub repository},
  howpublished = {\url{https://github.com/sporgj/nexus-code}},
  commit = {1181555484872f820a62798e557c3acc6c958d78}
}

@INPROCEEDINGS{segshare,
  author={Fuhry, Benny and Hirschoff, Lina and Koesnadi, Samuel and Kerschbaum, Florian},
  booktitle={2020 50th Annual IEEE/IFIP International Conference on Dependable Systems and Networks (DSN)},
  title={SeGShare: Secure Group File Sharing in the Cloud using Enclaves},
  year={2020},
  volume={},
  number={},
  pages={476-488},
  doi={10.1109/DSN48063.2020.00061}
}

@article{enclage,
author = {Sun, Yuanyuan and Wang, Sheng and Li, Huorong and Li, Feifei},
title = {Building Enclave-Native Storage Engines for Practical Encrypted Databases},
year = {2021},
issue_date = {February 2021},
publisher = {VLDB Endowment},
volume = {14},
number = {6},
issn = {2150-8097},
url = {https://doi.org/10.14778/3447689.3447705},
doi = {10.14778/3447689.3447705},
journal = {Proc. VLDB Endow.},
month = {feb},
pages = {1019–1032},
numpages = {14}
}

@inproceedings{orshuffle,
	author = {Sasy, Sajin and Johnson, Aaron and Goldberg, Ian},
	title = {Fast Fully Oblivious Compaction and Shuffling},
	year = {2022},
	isbn = {9781450394505},
	publisher = {Association for Computing Machinery},
	address = {New York, NY, USA},
	url = {https://doi.org/10.1145/3548606.3560603},
	doi = {10.1145/3548606.3560603},
	booktitle = {Proceedings of the 2022 ACM SIGSAC Conference on Computer and Communications Security},
	pages = {2565–2579},
	numpages = {15},
	location = {Los Angeles, CA, USA},
	series = {CCS '22}
}

\end{document}